\documentclass[10pt,journal,compsoc]{IEEEtran}
\ifCLASSOPTIONcompsoc
  \usepackage[nocompress]{cite}
\else
  \usepackage{cite}
\fi

\ifCLASSINFOpdf
\else
\fi

\usepackage[utf8]{inputenc} 
\usepackage[T1]{fontenc}    
\usepackage{hyperref}       
\usepackage{url}            
\usepackage{booktabs}       
\usepackage{nicefrac}       
\usepackage{microtype}      
\usepackage{multirow}

\usepackage{enumerate}
\usepackage{paralist}
\usepackage{caption}
\usepackage{subcaption}
\usepackage{graphicx}
\graphicspath{{figures/}}
\usepackage{mathtools,commath,amsmath,amssymb,amsfonts,bm}\interdisplaylinepenalty=2500
\usepackage{amsthm}
\usepackage{algpseudocode,algorithm}
\usepackage{tabularx,adjustbox}
\newcolumntype{L}[1]{>{\hsize=#1\hsize\raggedright\arraybackslash}X}%
\newcolumntype{R}[1]{>{\hsize=#1\hsize\raggedleft\arraybackslash}X}%
\newcolumntype{C}[1]{>{\hsize=#1\hsize\centering\arraybackslash}X}%

\usepackage{xspace}
\usepackage{todonotes}
\usepackage{breqn}
\clearpage
\extrafloats{100}
\RequirePackage{tcolorbox}
\tcbuselibrary{breakable}
\tcbset{parbox=false, left=1ex, right=1ex}

\def\vec#1{\mathchoice{\mbox{\boldmath$\displaystyle#1$}}
  {\mbox{\boldmath$\textstyle#1$}}
  {\mbox{\boldmath$\scriptstyle#1$}}
  {\mbox{\boldmath$\scriptscriptstyle#1$}}}

\newcommand{\Set}[1]{\mathcal{#1}}

\newcommand{\etal}{\textit{et~al.}\xspace}
\newcommand{\ie}{\textit{i.e.},\xspace}
\newcommand{\eg}{\textit{e.g.},\xspace}

\newtheorem{theorem}{Theorem}
\newtheorem{dfn}{Definition}
\newtheorem{exmp}{Example}[section]

\algnewcommand\algorithmicinput{\textbf{Input:}}
\algnewcommand\Input{\item[\algorithmicinput]}
\algnewcommand\algorithmicoutput{\textbf{Output:}}
\algnewcommand\Output{\item[\algorithmicoutput]}
\algnewcommand\algorithmictier{\textbf{Step:}}
\algnewcommand\Tier{\item[\algorithmictier]}

\usepackage{listings}
\lstdefinestyle{Lua}
{
  language         = {[5.1]Lua},
  basicstyle       = \linespread{0.9}\small\tt,
  showstringspaces = false,
  upquote          = true,
}

\usepackage[utf8]{inputenc}
\usepackage{tikz,pgfplots}
\usetikzlibrary{positioning}
\usetikzlibrary{calc}
\usetikzlibrary{fit}
\usetikzlibrary{decorations.text}
\usetikzlibrary{decorations.pathreplacing}
\usetikzlibrary{shapes.misc}
\usetikzlibrary{shapes.geometric,arrows}
\usetikzlibrary{decorations.pathmorphing}

\begin{document}
\title{How to Democratise and Protect AI: Fair and Differentially Private Decentralised Deep Learning}
\author{Lingjuan~Lyu,~\IEEEmembership{Member,~IEEE},
        Yitong~Li, Karthik~Nandakumar,~\IEEEmembership{Senior Member,~IEEE},
        Jiangshan~Yu, and~Xingjun~Ma
\IEEEcompsocitemizethanks{\IEEEcompsocthanksitem L. Lyu is with the Department of Computer Science, National University of Singapore. E-mail: lyulj@comp.nus.edu.sg.
\IEEEcompsocthanksitem Y. Li is with School of Computing and Information Systems, The University of Melbourne, Parkville, Australia, 3010. E-mail: yitongl4@student.unimelb.edu.au.
\IEEEcompsocthanksitem Karthik Nandakumar is with IBM Singapore Lab, 018983. E-mail: nkarthik@sg.ibm.com.
\IEEEcompsocthanksitem J. Yu is with Faculty of Information Technology,
Monash University, Clayton, Australia. E-mail: jiangshan.yu@monash.edu.
\IEEEcompsocthanksitem X. Ma is with School of Information Technology,
Deakin University, Geelong, Australia. E-mail: daniel.ma@deakin.edu.au.}
}

\markboth{IEEE TRANSACTIONS ON DEPENDABLE AND SECURE COMPUTING}%
{Lyu \MakeLowercase{\textit{\etal}}: How to Democratise and Protect AI: Blockchain-Enabled Decentralised Deep Learning with Differential Privacy}

\IEEEtitleabstractindextext{
\begin{abstract}
This paper firstly considers the research problem of fairness in collaborative deep learning, while ensuring privacy. A novel reputation system is proposed through digital tokens and local credibility to ensure fairness, in combination with differential privacy to guarantee privacy. In particular, we build a fair and differentially private decentralised deep learning framework called FDPDDL, which enables parties to derive more accurate local models in a fair and private manner by using our developed two-stage scheme: during the initialisation stage, artificial samples generated by Differentially Private Generative Adversarial Network (DPGAN) are used to mutually benchmark the local credibility of each party and generate initial tokens; during the update stage, Differentially Private SGD (DPSGD) is used to facilitate collaborative privacy-preserving deep learning, and local credibility and tokens of each party are updated according to the quality and quantity of individually released gradients. Experimental results on benchmark datasets under three realistic settings demonstrate that FDPDDL achieves high fairness, yields comparable accuracy to the centralised and distributed frameworks, and delivers better accuracy than the standalone framework.
\end{abstract}

}

\maketitle

\IEEEdisplaynontitleabstractindextext

\begin{figure*}[!htp]
\centering
        \begin{subfigure}[ht]{0.22\textwidth}
        \centering \includegraphics[width=3cm,height=2.3cm]{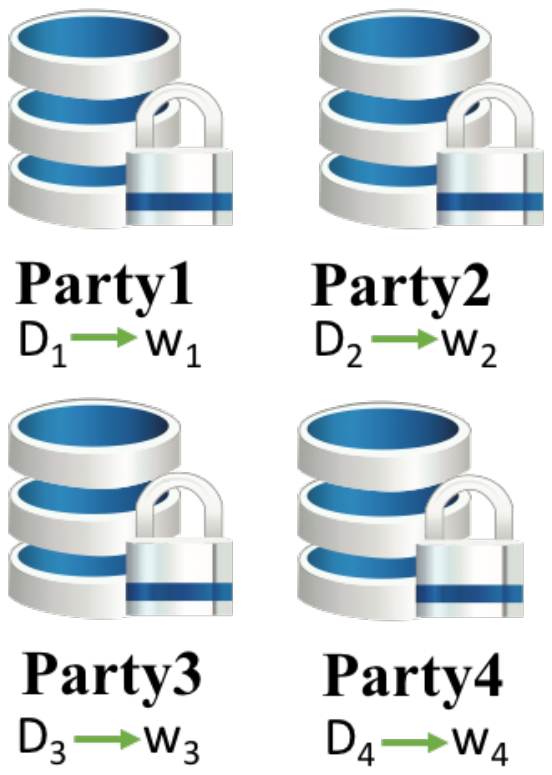}
		\subcaption{\label{fig:standalone} Standalone.}
        \end{subfigure}
        \begin{subfigure}[ht]{0.22\textwidth}
        \centering
                \includegraphics[width=3cm,height=2.3cm]{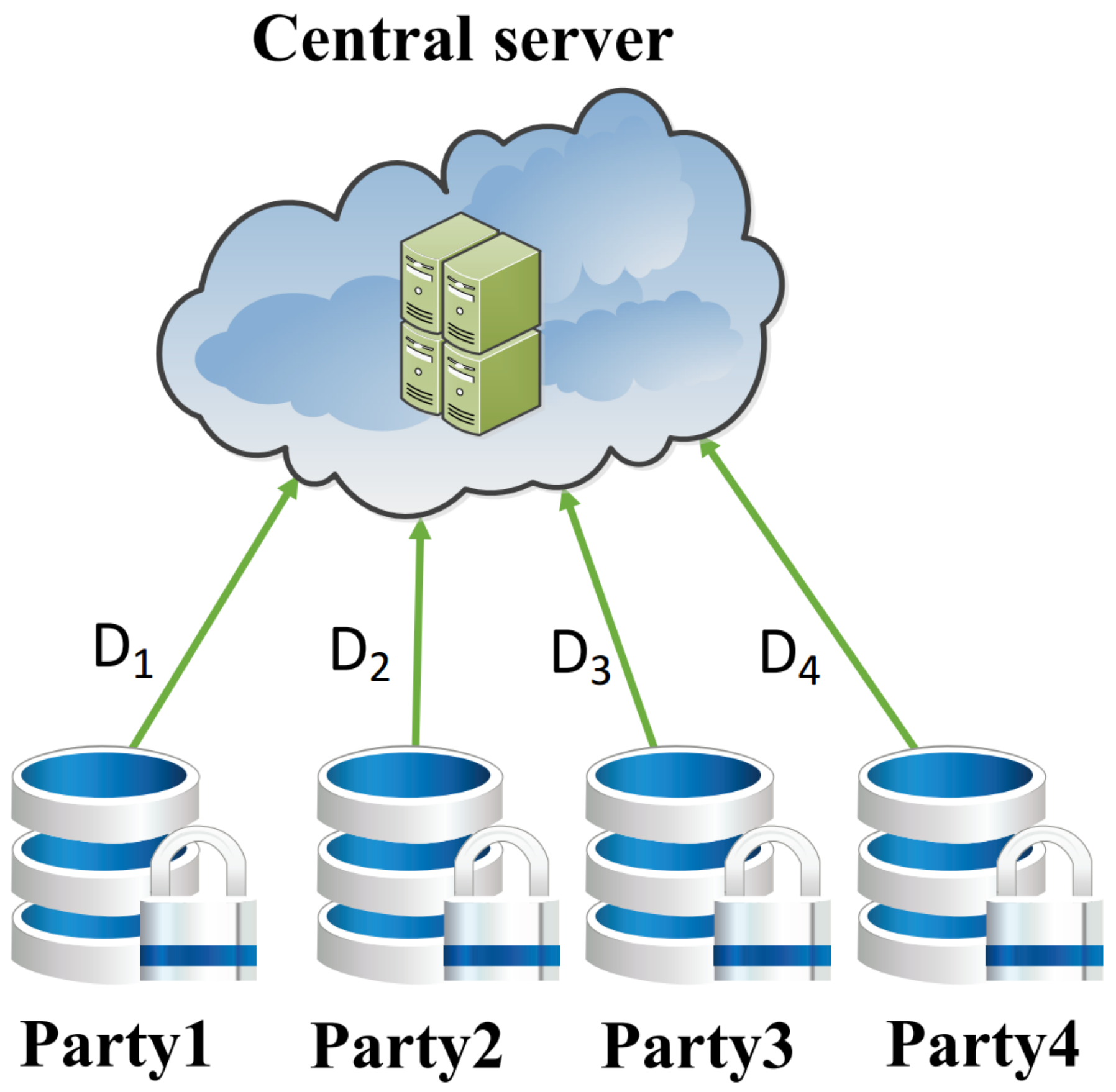}
                \subcaption{\label{fig:centralised} Centralised.}
        \end{subfigure}
        ~
         \begin{subfigure}[ht]{0.22\textwidth}
        \centering
                \includegraphics[width=3cm,height=2.3cm]{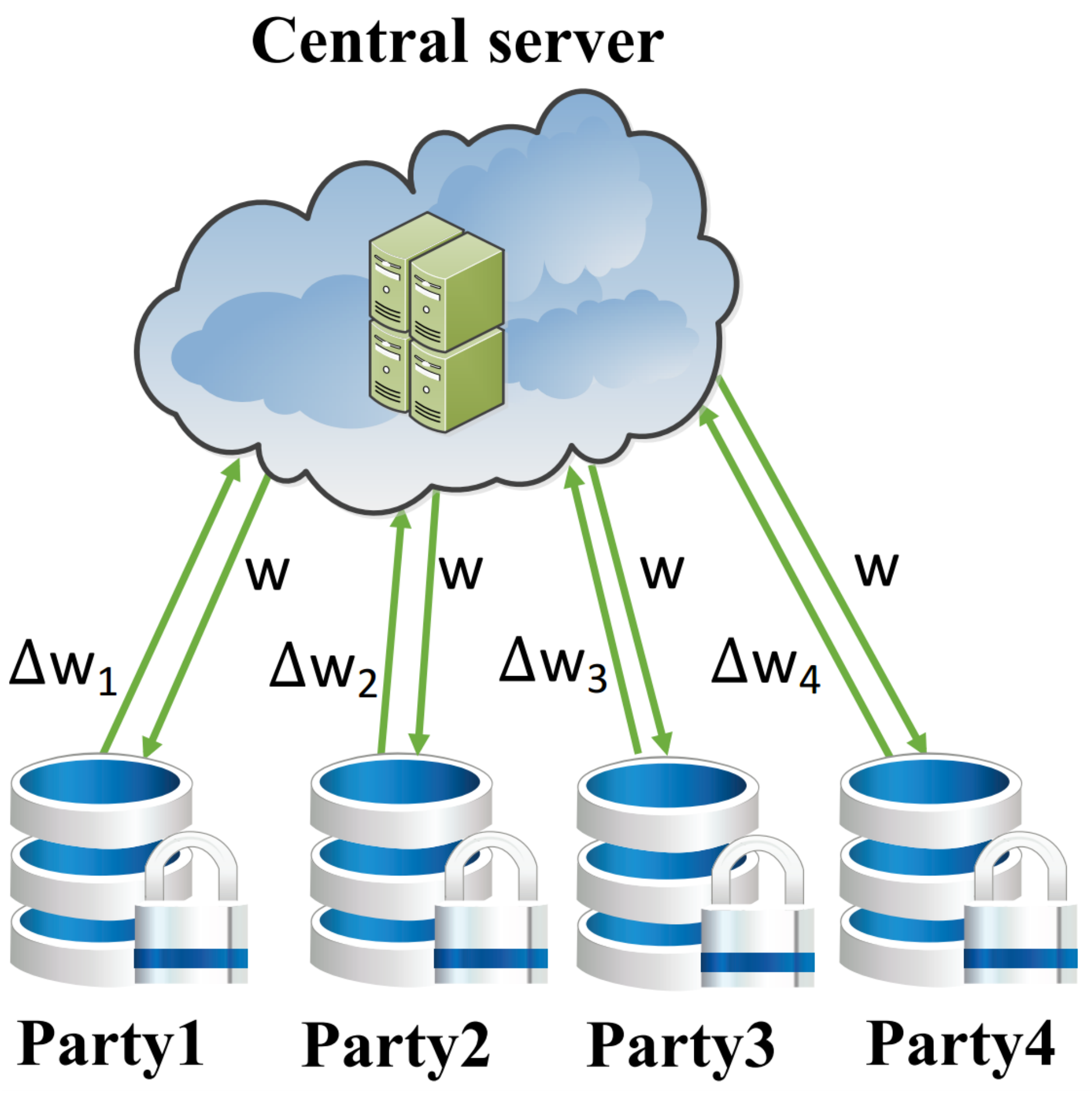}
                \subcaption{\label{fig:distributed} Distributed.}
        \end{subfigure}
        ~
        \begin{subfigure}[ht]{0.22\textwidth}
        \centering
                \includegraphics[width=3cm,height=2.3cm]{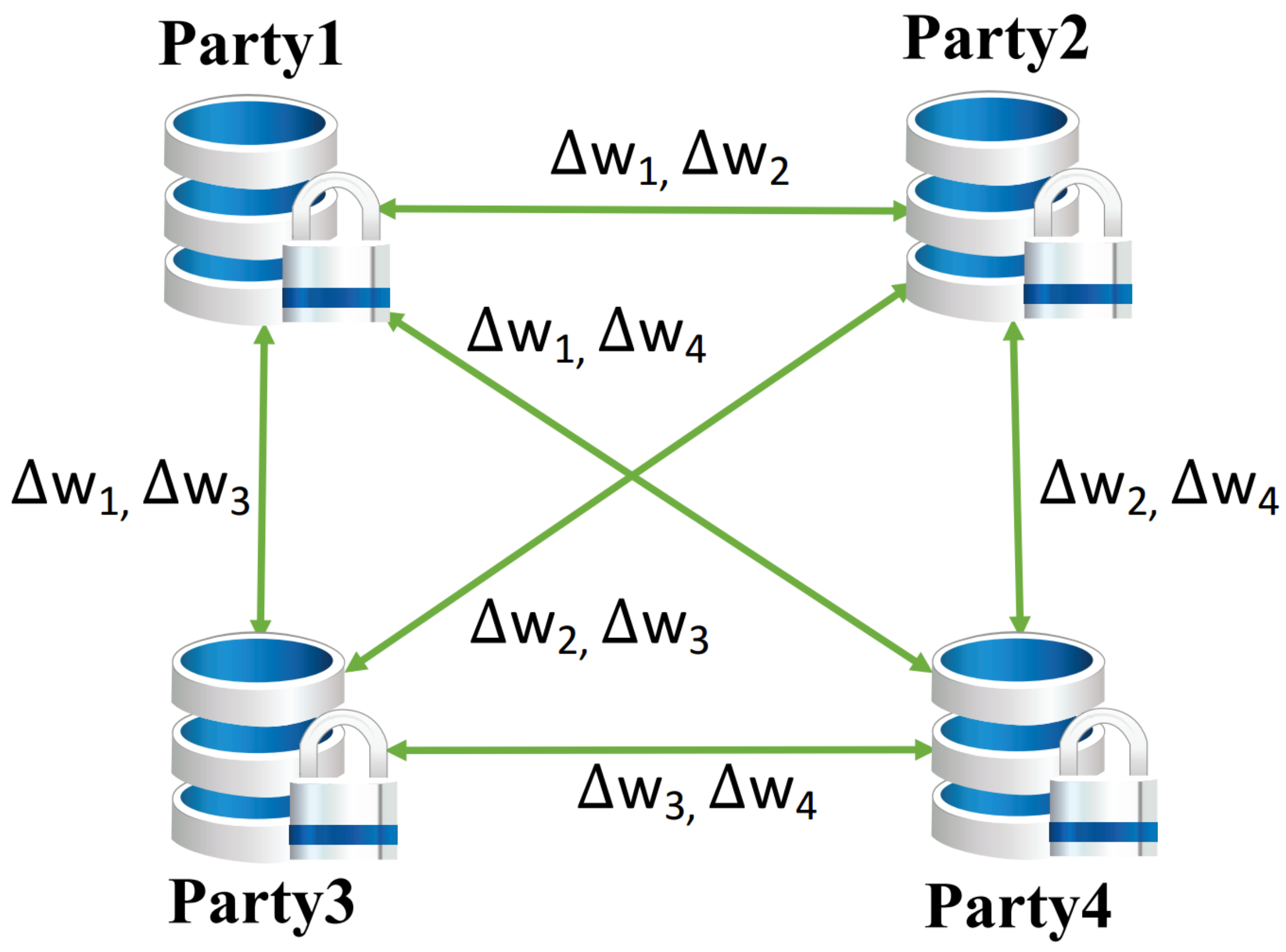}
                \subcaption{\label{fig:decentralised} Decentralised.}
        \end{subfigure}
        \caption{Different deep learning frameworks.}
\label{fig:model}
\vspace{-3mm}
\end{figure*}

\IEEEpeerreviewmaketitle
\begin{sloppypar}
\IEEEraisesectionheading{\section{Introduction}\label{sec:introduction}}
\IEEEPARstart{I}{n} real world, many practical applications would benefit from large-scale deep learning across sensitive datasets owned by different parties, thus data sharing and analysis across parties are of paramount importance to accelerate scientific discovery, facilitate quality improvement initiatives, speed up hypothesis testing, and boost accuracy towards higher level, especially when there are not enough local examples to test a hypothesis~\cite{ohno2012idash}.
This trend is motivated by the fact that the data from a single organization may be very homogeneous, ending up with an unsatisfactory model that fails to generalise to other data, as shown in Fig.~\ref{fig:standalone}.
Therefore, there is much demand to train a global model by a central server on the combined data collected from independent parties to ensure sufficient statistical power to test hypotheses (Fig.~\ref{fig:centralised}).
On the other hand, deep learning can be performed in a collaborative manner, where a parameter server is required to maintain the latest parameters available to all parties (Fig.~\ref{fig:distributed}). However, such central server-based learning framework suffers from the following weaknesses:
\begin{compactitem}
\item \textbf{Untrusted server}. Due to privacy issue, a party may not trust a central server~\cite{mcconaghy2016bigchaindb}, thus reluctant to transfer either data or model parameters to the server.

\item \textbf{Single-point-of-failure}. Once the central server is shut down, the whole network stops working. Moreover, if the central server is attacked, the entire network is under the risk of being compromised~\cite{mcconaghy2016bigchaindb}.

\item \textbf{Malicious attack}. The data being disseminated is mutable. An attacker could arbitrarily change its local model without being detected, and no audit trail is available to identify such malicious behaviour.

\item \textbf{Lack of fairness and vulnerable to free-riders}: Existing frameworks consider that all parties contribute equally. This is typically impractical due to the data quality and quantity of each party. It is thus unfair that at the end of the collaboration, all parties get access to the same global model regardless of their contributions. In an extreme case, even the free-riders could successfully join the system, and enjoy the system's global model for free. The lack of fairness might discourage collaboration among parties. 

\end{compactitem}

To address the first two issues, we make use of Blockchain to provide a fully decentralised framework, \ie each participant \textbf{does not} trust any third party or any other participants, as illustrated in Fig.~\ref{fig:decentralised}. 
In particular, we explicitly study two types of malicious parties, including free-rider without any data and the GAN attacker. We \emph{claim these two malicious parties considered all belong to the category of ``non-credible'' parties, but not all ``non-credible'' parties are malicious}, a ``non-credible'' party might follow the protocol honestly, but may have limited data or totally different data distribution from the majority party, thus it is still reasonable for all the other parties to give low local credibility to this ``non-credible'' party, or even isolate it.

However, the last two issues are yet to be solved, and collaboration might be significantly hindered due to privacy and confidentiality restrictions. To overcome these problems, we are inspired to develop a decentralised collaborative learning framework that respects collaborative fairness, data privacy and utility at the same time, so as to encourage more parties to collaborate. \textbf{Our main contributions} are summarised as follows: 

\begin{itemize}
\item We initiate the research problem of collaborative fairness in collaborative learning, and propose a fair and private framework, named Fair and Differentially Private Decentralised Deep Learning (FDPDDL). 
\item To address fairness and privacy problem, we propose a two-stage scheme, which is realised by two algorithms respectively, \ie i) local credibility and tokens initialisation, and ii) local credibility and tokens update. In particular, we build a novel reputation system, which reflects the relative contribution of each party, thus ensuring fairness; we use \emph{Differentially Private GAN} (DPGAN) in the initialisation stage and \emph{Differentially Private Stochastic Gradient Descent} (DPSGD) in the update stage to mitigate privacy leakage; 

\item Our framework provides a viable solution to detect and isolate free-riders and GAN attacker both before and during the collaborative learning process;

\item We evaluate our framework on several benchmark datasets under three realistic settings.
Extensive experiments demonstrate that FDPDDL achieves high fairness, delivers comparable accuracy to the centralised and distributed deep learning frameworks, and outperforms the standalone deep learning framework, confirming the superiority of FDPDDL. 
\end{itemize}

\section{Related Work and Preliminaries}
\label{sec:Related_work}
This section firstly reviews the relevant deep learning frameworks, and the privacy and fairness issues in deep learning. Secondly, the relevant techniques used in this paper are introduced, including differential privacy and Blockchain. 

\subsection{Deep Learning Frameworks}
In general, deep learning frameworks fall into the following four categories: 

\textit{\textbf{Standalone deep learning}}: participants individually train standalone models on their training data without any collaboration, as shown in Fig.~\ref{fig:standalone}. 

\textit{\textbf{Centralised deep learning}}: Centralised deep learning forces multiple participants to pool their data into a centralised server to train a global model on the combined data, as depicted in Fig.~\ref{fig:centralised}.

\textit{\textbf{Distributed deep learning}}: Shokri \etal~\cite{shokri2015privacy} firstly introduced the concept of \emph{Distributed Selective Stochastic Gradient Descent} (DSSGD) for distributed deep learning. It allows each party to keep local model private while iteratively update its model by integrating differentially private gradients from other parties via a parameter server, as illustrated in Fig.~\ref{fig:distributed}. The communication cost, within each round of parameter update, is addressed by only sharing a fraction (\eg 1\%-10\%) of local model gradients that have values larger than a certain threshold or those gradients with the largest absolute values. 

Federated learning is a special case of distributed deep learning, which is tailored to deal with Non-IID, unbalanced and massively distributed data in mobile application~\cite{mcmahan2017communication,bonawitz2017practical,mcmahan2018learning}. The goal is to train a shared global model while leaving training data on users' mobile phones. Mobile phones with relatively powerful and fast processors (including GPUs) are required to download the current model, compute updates by performing local computation, then send local model updates to the trusted Google Cloud server in each communication round. 

\textit{\textbf{Decentralised deep learning}}: 
Decentralised framework is much different than server-based framework, in the sense that it is purely decentralised without relying on any central servers, as exemplified in Fig.~\ref{fig:decentralised}.
The first decentralised machine learning model is ModelChain~\cite{Kuo2016ModelChain}, which applies Blockchain technology to machine learning by incorporating the idea of boosting, \ie samples that are more difficult to classify are more likely to improve the model significantly. The follow-up work~\cite{zhu2018blockchain,kim2019blockchained,kang2019incentive} integrated blockchain into deep learning. For example, Kang \etal~\cite{kang2019incentive} proposed an effective incentive mechanism to motivate high-reputation mobile devices with high-quality data to participate in model learning, but they overlooked the privacy issues.

It should be noted that in both distributed framework and decentralised framework, parties are all involved in the iterative process of building a global or consensus model, hence we call them \emph{collaborative deep learning frameworks}. A succinct comparison among different deep learning frameworks is provided in Table~\ref{tbl:cc}.

\begin{table*}[ht]
  \centering\caption{\label{tbl:cc}Feature comparison of different deep learning frameworks.}
  \begin{tabularx}{\textwidth}{p{3.8cm}cccccc}
  \toprule
   Deep learning frameworks & Standalone & Centralised & Distributed~\cite{shokri2015privacy,mcmahan2017communication,mohassel2017secureml} & Decentralised~\cite{Kuo2016ModelChain} & Decentralised (ours)\\
  \midrule
  Architecture & Fig.~\ref{fig:standalone} & Fig.~\ref{fig:centralised} & Fig.~\ref{fig:distributed} & Fig.~\ref{fig:decentralised} & Fig.~\ref{fig:decentralised} \\
  Global/Consensus model
    & No
    & Yes
    & Yes 
    & Yes
    & No \\
  Local models
    & Yes
    & No
    & Yes
    & Yes
    & Yes \\
  Fairness
    & NA
    & NA
    & No
    & No
    & Yes \\
  ``non-credible'' party detection
    & NA
    & No
    & No
    & No
    & Yes \\
  \bottomrule
  \end{tabularx}
\end{table*}

\subsection{Privacy-preserving Deep Learning}
\textit{\textbf{Privacy-preserving Centralised deep learning}}: Centralised model is very effective, however it is not privacy-preserving since the central server has direct access to all sensitive information. Shokri \etal~\cite{shokri2015privacy} pointed out that centralised deep learning poses serious privacy threats, including (i) all the sensitive training data are exposed to a susceptible third party who can permanently keep the collected data; (ii) data owners have no control over the learning objective or the knowledge of what can be inferred from their data; (iii) the learned model is not directly available to data owners. 

\textit{\textbf{Privacy-preserving Distributed deep learning}}: 
Distributed deep learning generally suffers from the common issue of privacy leakage from the shared gradients.
As demonstrated in~\cite{aono2018privacy}, even a small proportion of local gradients can reveal certain amount of local data information. In the case of a local network with only one neuron, the server can extract local data with non-negligible probability. Even for complex neural networks trained with regularisation, the gradients can still expose certain label information of local data~\cite{aono2018privacy}. Moreover, if a party turns out to be malicious, it can easily sabotage the learning process (\eg by spoofing random data samples) or violate some of the privacy requirements by inferring information about the victim party's private data, which the attacker is not supposed to know. Hitaj \etal~\cite{hitaj2017deep} devised an active inference attack on deep neural networks in a collaborative setting, which is referred to as \emph{Generative Adversarial Networks} (GAN) attack. It exploits the real-time nature of the learning process that allows the adversarial party to train a GAN that generates prototypical samples of the targeted training data that was meant to be private and the generated samples are intended to come from the same distribution as the training data. The malicious party is able to attack other parties successfully as long as the global model is under the process of learning. GAN attack makes the distributed setting even more undesirable, as in centralised learning only the server may pose privacy threat, but in distributed learning, any party can violate the privacy of any other parties in the system, even without involving the server~\cite{hitaj2017deep}. It is worth noting that GAN attack succeeds only if the following three conditions are held: (i) the adversary has knowledge of labels of the victim party; (ii) class distributions of the adversary and the victim party are non-independent and identically distributed (Non-IID); (iii) the victim party is not secured by any privacy protection mechanism or it adopts per-parameter privacy in DSSGD which results in meaningless privacy.

To tackle with privacy issue, secure multiparty computation (SMC) has been used to build privacy-preserving neural networks in a distributed manner. For example, SecureML~\cite{mohassel2017secureml} allows clients to distribute their private training data among two non-colluding servers during the setup phase; these two servers then employ SMC to train a global model on the clients' encrypted joint data. In general, SMC techniques achieve a high level of privacy and accuracy, at the expense of high computational and communication overhead for participants, thereby doing a disservice to attracting participation. 
Alternatively, Shokri \etal~\cite{shokri2015privacy} perturbed the shared local model gradients by adding noise to satisfy differential privacy. However, their privacy bounds are given per-parameter, the large number of parameters prevents the technique from providing a meaningful privacy guarantee. 

In federated learning, to protect individual model updates from the adversarial server who might scrutinize individual updates, instead of using differential privacy as in~\cite{shokri2015privacy}, Bonawitz \etal~\cite{bonawitz2017practical} proposed a secure and failure-robust protocol based on SMC to securely aggregate local model updates as the weighted average to update the global model on the server. Another more efficient method is to borrow differential privacy to conceal user participation, as demonstrated by McMahan \etal~\cite{mcmahan2018learning}. However, it requires a large number of users (on the order of thousands) to ensure model convergence and an acceptable trade-off between privacy and utility. Moreover, the default trusted Google server is entitled to see all users' updates in the clear, aggregate these updates and add noise to the aggregation, hence their scheme is even weaker than DSSGD when the server is untrusted. 

Overall, all the current distributed deep learning frameworks need to be coordinated by a central server, thus falling under the umbrella of server-based frameworks.

\textit{\textbf{Privacy-preserving Decentralised deep learning}}: 
The first decentralised machine learning model, \ie ModelChain, stated that privacy is preserved by exchanging zero patient data, however, the exchanged model-level information can still largely leak local data information~\cite{aono2018privacy}. Furthermore, the proposed logic for ModelChain is reasonable only if all the participants are honest. More recently, Kim \etal~\cite{kim2019blockchained} proposed blockchain-based privacy preserving deep learning and utilized a consensus mechanism to verify local model updates. Zhu \etal~\cite{zhu2018blockchain} provided a proof-of-concept for managing security issues in federated learning systems via blockchain technology. However, none of these works considered the fairness problem in collaborative learning. 

\subsection{Fairness in Collaborative Deep Learning}
There has been a long line of work studying fairness in machine learning, however, to the best of our knowledge, existing research on fairness mostly focuses on the protection of some specific attributes, or aim to reduce the variance of the accuracy distribution across participants~\cite{cummings2019compatibility,jagielski2019differentially}, while none of the previous works addressed the problem of collaborative fairness in collaborative learning. 

Overall, all the current collaborative deep learning frameworks focus on how to learn a global model or consensus model with higher accuracy than standalone models, while losing the ability to verify the contribution of individual participant, because participants can access the same global model or consensus model no matter how differently they contribute. In extreme cases, there may exist free-riders in the collaborative learning system, who aim to benefit from the global model, but do not want to contribute any real information. For clarity, we give a concrete example in Example.~\ref{exmp:malicious} to showcase how a free-rider party $C$ (no data or model in particular) can also obtain the global model even if it fails to make any practical contributions to the global learning process.
\begin{exmp}\label{exmp:malicious}
Suppose that three parties A, B, and C are involved in server-based deep learning:
\begin{compactitem}
  \item A (honest) has data $D_A$ and Model $M_A$ with accuracy $90\%$.

  \item B (honest) has data $D_B$ and Model $M_B$ with accuracy $80\%$.

  \item C (free-rider) has no data or model to start with.
\end{compactitem}
Suppose that whenever it is C's turn to upload the gradients, it always uploads random or carefully crafted gradients, so that the global learning process will not be affected. Finally, all three participants have access to the same global model.
\end{exmp}

\emph{This lack-of-fairness is an essential problem that persists in all the existing collaborative learning frameworks but has been overlooked by far}. Lack of fairness can be an obstacle for widespread the adoption of collaborative learning as a new type of powerful learning platform. 
On the other hand, this fairness issue can be addressed by swinging existing learning frameworks to the other extreme that openly publishes all the gradients. In that case, fairness might be achieved but at the cost of privacy, which is highly undesirable in collaborative deep learning. 

\subsection{Differential Privacy}
\begin{dfn}\label{def:diff-pri}
A randomised algorithm $\mathcal{A}$ satisfies $(\epsilon,\delta)$-approximate Differential Privacy (DP) if 
\begin{equation}\label{eq:diff-pri2}
	\Pr\{\mathcal{A}(D_1)\in\Set{S}\} \leq e^\epsilon\Pr\{\mathcal{A}(D_2)\in\Set{S}\} + \delta \, ,
\end{equation}
for all set $\Set{S}\subseteq\mathrm{range}(\mathcal{A})$, and all pairs of datasets $D_1$, $D_2$, where $D_1$ can be obtained from $D_2$ by adding \emph{or} removing one tuple. Further, if $\delta=0$, we say $\mathcal{A}$ preserves $\epsilon$-differential privacy.
\end{dfn}

Unlike the previous empirical criterion for privacy~\cite{lyu2017privacy}, differential privacy is based on a solid theoretical foundation~\cite{dwork2014algorithmic}. The formal definition of DP has two parameters: i) privacy budget $\epsilon$ measures the incurred privacy leakage -- lower $\epsilon$ means less information leakage and higher privacy guarantee; ii) $\delta$ bounds the probability that the privacy loss exceeds $\epsilon$, with the recommended value $\delta \ll 1/N$, where $N$ is the number of training examples. The values of $(\epsilon,\delta)$ accumulate as the algorithm repeatedly accesses the private data~\cite{abadi2016deep}. 

\begin{theorem}
\label{Theorem_dpcom}
\cite[Theorem 3.16.]{dwork2014algorithmic} Composition for ($\epsilon,\delta$)-differential privacy (the epsilons and the deltas add up): the composition of $k$ differentially private mechanisms is $({\textstyle\sum}_i \epsilon_i, {\textstyle\sum}_i \delta_i)$-differentially private, where for any $1 \leq i \leq k$, the $i$-th mechanism is $(\epsilon_i, \delta_i)$-differentially private.
\end{theorem}

\subsection{Blockchain Technology}
Blockchain was first proposed as a proof-of-work consensus protocol implementation of peer-to-peer timestamp server on a decentralised basis in the Bitcoin crypto-currency~\cite{nakamoto2019bitcoin}. As a new form of a distributed database, it can store arbitrary data in the transaction metadata. Specifically, an electronic coin (\eg Bitcoin) is defined as a chain of transactions. A block contains multiple transactions to be verified, and the blocks are chained as blockchain using hash functions to achieve the timestamp feature.
Such Blockchain-based distributed database is known as Blockchain 2.0, including technologies such as smart properties (the properties with blockchain-controlled ownership) and smart contracts (programs that manage smart properties)~\cite{wood2014ethereum}. In the context of a distributed database, smart properties are data entries, and smart contracts are stored procedures. Therefore, our FDPDDL can be implemented using Blockchain 2.0 technologies, where the transaction metadata is utilised to disseminate local DPGAN samples or model gradients among parties, all the upload and download transactions are recorded immutably on the blockchain, and algorithms like tokens/credibility assignments are done using smart contracts, which make all the transactions among all parties fully visible. Compared with current server-based frameworks, the peer-to-peer architecture of Blockchain allows each party to remain modular while interoperating with other parties. In addition, instead of ceding control to the central server, Blockchain enhances security by avoiding single-point-of-failure, each party in the Blockchain system has control about how its data should be accessed, hence obeying the institutional policies. 

\section{FDPDDL Framework}
\label{sec:FDPDDL}
This section details our proposed Fair and Differentially Private Decentralised Deep Learning (FDPDDL) framework, including the main focuses of FDPDDL, and an investigation on the Blockchain as the decentralised architecture for FDPDDL. 
For the readers' convenience, Table~\ref{tbl:symbs}
contains a list of notations used throughout the paper.
 
 \begin{table}[ht]
    \centering\caption{\label{tbl:symbs}Table of notations.}
    \begin{tabularx}{\linewidth}{cX}
    \toprule
    Symbol & Meaning \\
    \midrule
    $D_i, V_i, M_i$ & local training data, validation data and standalone model of party $i$\\
    $p_i$, $d_i$ & tokens and gradients download budget of party $i$ \\
    $c_i^j$ & local credibility of party $j$ given by party $i$ based on the usefulness of party $j$ to party $i$\\
    $u_i$ & number of uploaded DPGAN samples or gradients by party $i$\\
    $d_j^i$ & number of gradients of party $j$ downloaded by party $i$\\
    $\lambda_j$ & sharing level of party $j$ \\
    $\Delta \vec{w}_j$ & gradients of party $j$ \\
    $\Delta (\vec{w}_j^i)^S$ & selected gradients of party $j$ sent to party $i$\\
    $\vec{w}_i$ & parameter of party $i$ at previous communication round\\
    $\vec{w}_i'$ & updated parameter of party $i$ at current round by combining all parties' selected gradients\\ 
    ${\vec{w}_i^j}'$ & temporary parameter of party $i$ by removing party $j$'s gradients $\Delta (\vec{w}_j^i)^S$ from $\vec{w}_i'$ to update $c_i^j$\\
    $acc$ & validation accuracy of party $i$\\
    $acc_j$ & validation accuracy of party $i$ by excluding party $j$'s $\Delta (\vec{w}_j^i)^S$\\
    $n$ & number of participating parties \\
    $c_{th}$ & lower bound of the credibility threshold  agreed by the majority party \\
    $C$ & credible party set with local credibility above $c_{th}$ \\ 
    $m_j$ & number of matches between majority labels and party $j$'s predicted labels  \\
    $e_i$ & gap between download budget $d_i$ and current downloads ${\textstyle\sum}_{j \in C \setminus i} d_j^i$ of party $i$\\
    $r_j^i$ & extra gradients of party $j$ that can be provided to $i$\\
    $L_i$ & parties in $C$ that can provide additional gradients to party $i$\\
    $(sk'_i, pk'_i)$ &  party $i$'s key pair for signing and verification, respectively \\
    $fsk$ & fresh symmetric encryption key used in the hybrid cryptosystem \\
    $(sk_i,pk_i)$ & party $i$'s key pair for decryption and encryption used in the hybrid cryptosystem\\
    \bottomrule
    \end{tabularx}
\vspace{-3mm}
\end{table}

\subsection{Main focuses of FDPDDL}
\label{subsec:focuses}
\textbf{Privacy:} In FDPDDL, we assume parties do not trust each other or any third server. To remove the deterrents for parties to collaborate, instead of publishing all the original data or model parameters, each party leverages DPGAN to publish differentially private samples for mutual evaluation during the initialisation stage, and publishes differentially private gradients during the update stage.

\textbf{Fairness:} 
The basic idea of fairness is motivated by the fact that the party who contributes more to other parties should be given a higher local credibility and rewarded with a better performing final model than a low-contribution party. To ensure fairness, we build a reputation system through digital tokens and local credibility. Each party in the system participates in the evaluation of the usefulness of other parties, and requests more samples or gradients from parties with higher local credibility. In this way, participants are motivated to release more in order to earn more tokens, which can be used to download gradients from other parties. For example, if a local model has 100K parameters, the participant with sharing level $\lambda_j$ of 0.1 can at most publish 10\%, \ie 10K gradients in a privacy-preserving manner and be rewarded 10K tokens. If any of the other participants want to download gradients, they need to pay some tokens. Uploading more samples or gradients gives a participant more tokens and using these tokens this participant can download more gradients published by others. This is the incentive for publishing more, as long as it is within the limits of privacy. Similarly, downloading more gradients consumes more tokens. Fairness is achieved by rewarding each party as per its relative contribution to other parties during download and upload processes as follows:

\begin{itemize}
\item  \textbf{Download as per local credibility}: Since one party might contribute differently to different parties, the credibility of one party might be different from the perspective of different parties, therefore, each party $i$ should keep a local credibility list by sorting all parties as per their local credibilities in descending order, which is known only by party $i$. The higher the local credibility of party $j$ in party $i$'s credibility list, the more likely party $i$ will download gradients from party $j$, and more tokens will be rewarded to party $j$.

\item  \textbf{Upload as per request and sharing level}: Once one party receives download request (demand for the number of gradients), how many gradients will be uploaded by the requested party depends on both the download request and the sharing level $\lambda_j$ of the requested party.
\end{itemize}

By enforcing fairness, our FDPDDL allows parties to (i) independently converge to different parameters; (ii) critically avoid overfitting their parameters to a single party's local training data. Once multiple local models are collaboratively trained, each party can independently evaluate its model on the unseen data, without interacting with other parties. 

\subsection{Blockchain investigation for FDPDDL}
For investigation of Blockchain, we first formulate the notion of ``digital token'' as a currency for transaction. Second, we make use of blockchain to record and supervise data exchange in a distributed manner that is robust, fair, and transparent. In particular, the differentially private samples or gradients will be traded using digital tokens, and recorded as transactions in the blockchain. Tokens can be consumed by downloading or be earned by uploading differentially private samples or gradients. In this way, we guarantee the fair exchange among participants.

Depending on the application scenario, the blockchain in FDPDDL can be either a consortium blockchain that requires permissions to participant in the system, such as Hyperledger Fabric \cite{hyperledger-fabric}, or
a permissionless blockchain which anyone can join at any time, such as
Ethereum \cite{ethereum}.

\subsubsection{Genesis Block}\label{sec:benchmarking}
In our blockchain, the first block, \ie the genesis block, initialises the system. The genesis
block also records the verification key $pk'_i$ of party $i$'s signing
key $sk'_i$. Whenever a new party joins in the network or the existing
party adds new data during training, initialisation will be restarted
and a new block will be added to the blockchain to update the relevant
data. Meanwhile, based on the Blockchain mechanism, we do not need to
deal with the party departure. When a party leaves the private
blockchain network, other parties just need to remove it from their
local credibility lists.

To initialise a genesis block, we propose Algorithm~\ref{Algorithm:benchmarking} to initialise local credibility and tokens, where participants contribute their artificial samples in a privacy-preserving manner, then mutually evaluate the quality of each other's samples, and gain reward in the form of tokens. 
All the released samples are authenticated through a digital signature scheme, where the public key $pk_i$ of party $i$ is advertised together with the signed samples. This key will later be included in the genesis block, and the corresponding signing key $sk'_i$ will be used to claim the associated reward. The agreed reward, in the form of tokens of each party, together with their verification key $pk'_i$, will be recorded in the genesis block through an initial blockchain consensus process, which is specific to each blockchain.

\subsubsection{Operation Block}
\label{sec:credit_update}
After initialisation, the event of trading (\ie purchasing)
differentially private gradients are recorded as
transactions in the blocks. To order gradients from a
party $j$, party $i$ needs to create a purchase order, as a
transaction, and record it in a block. The transaction includes the
tokens party $i$ is willing to pay for a specified number of gradients, and party $i$'s public key $pk_i$ that will be used later by party $j$ for encryption purpose. Considering the released gradients are of high dimensionality, the standard hybrid cryptosystem can be used to take advantage of the efficiency of the symmetric-key cryptosystem --- a freshly generated symmetric key $fsk$ is used to encrypt gradients, while $pk_i$ is used to encrypt $fsk$. In this way, we minimize the required computational cost incurred by asymmetric key based encryption. Once the order is placed in the blockchain, party $j$ agrees and completes the order in two steps: (1) party $j$ encrypts the selected gradients with $fsk$, and sends both the encrypted gradients and encrypted $fsk$ to a public accessible storage; (2) party $j$ creates a transaction that contains the hash value of the encrypted gradients, together with a pointer to the
transaction containing party $i$'s request. Once this transaction is included in the blockchain, the agreed tokens will be transferred from party $i$ to party $j$ automatically through blockchain. Note that if party $i$ wants to maliciously denial the fact that party $j$ has honestly shared gradients, party $j$ can reveal the provided gradients to other participants for verification. Since both $pk_i$ and the hash value of the encrypted gradients are recorded in the blockchain, with the gradients revealed by party $j$, anyone can verify that whether party $j$ has provided the requested data to party $i$, and party $i$ will be punished through a special transaction in the blockchain. Similarly, thanks to the transparent blockchain, if party $j$ is misbehaved, it will be detected and punished in a similar way.

\section{FDPDDL Realization}
This section details the two-stage realization in FDPDDL: local credibility and tokens initialisation and local credibility and tokens update, as shown in Fig.~\ref{fig:FDPDDL_flow}.

\def\dpgan#1{
\begin{scope}[shift={#1}]
    \draw [rounded corners,ultra thick] (-.6,-.6) rectangle (0.6, 0.6);
    
    \draw [fill=white, thick] (-0.45,0.15) circle (0.05);
    \draw [fill=white, thick] (-0.15,0.15) circle (0.05);
    \draw [fill=white, thick] (0.15,0.15) circle (0.05);
    \draw [fill=white, thick] (0.45,0.15) circle (0.05);

    \draw [fill=white, thick] (-0.45,-0.15) circle (0.05);
    \draw [fill=white, thick] (-0.15,-0.15) circle (0.05);
    \draw [fill=white, thick] (0.15,-0.15) circle (0.05);
    \draw [fill=white, thick] (0.45,-0.15) circle (0.05);

    \draw [fill=white, thick] (-0.15,0.4) circle (0.05);
    \draw [fill=white, thick] (0.15,0.4) circle (0.05);
    
    \draw [fill=white, thick] (0,-0.4) circle (0.05);
    
    \draw[] (-0.15,0.35)--(-0.45,0.2);
    \draw[] (-0.15,0.35)--(-0.15,0.2);
    \draw[] (-0.15,0.35)--(0.15,0.2);
    \draw[] (-0.15,0.35)--(0.45,0.2);
    
    \draw[] (0.15,0.35)--(-0.45,0.2);
    \draw[] (0.15,0.35)--(-0.15,0.2);
    \draw[] (0.15,0.35)--(0.15,0.2);
    \draw[] (0.15,0.35)--(0.45,0.2);
    
    \draw[] (0,-0.35)--(-0.45,-0.2);
    \draw[] (0,-0.35)--(-0.15,-0.2);
    \draw[] (0,-0.35)--(0.15,-0.2);
    \draw[] (0,-0.35)--(0.45,-0.2);
    
    \draw[decorate,decoration={snake,amplitude=.4mm,segment length=2mm}] (-0.5, 0)--(0.5,0);
    
\end{scope}
}
\begin{figure*}[!htp]
\centering 
\resizebox{0.8\textwidth}{!}{\begin{tikzpicture}[line width=0.03cm]
    \node[align=center,minimum size=0.3cm] at (-2, 2) {Party 1 \\ $D_1$,$M_1$};
    \node[align=center,minimum size=0.3cm] at (2, 2) {Party 2 \\ $D_2$,$M_2$};
    \node[align=center,minimum size=0.3cm] at (-2, -2) {Party 3 \\ $D_3$,$M_3$};
    \node[align=center,minimum size=0.3cm] at (2, -2) {Party 4 \\ $D_4$,$M_4$};
    
    \draw[draw=cyan] (-2, 2) circle (0.7);
    \draw[draw=cyan] (2, 2) circle (0.7);
    \draw[draw=cyan] (-2, -2) circle (0.7);
    \draw[draw=cyan] (2, -2) circle (0.7);
    
    \node[align=center,minimum size=0.3cm] at (0, 2.3) {$SD_1$};
    \draw[->] (-1.3, 2.1)--(1.3, 2.1);
    \node[align=center,minimum size=0.3cm] at (0, 1.7) {$M_2(SD_1)$};
    \draw[->] (1.3, 1.9)--(-1.3, 1.9);
    
    \node[align=center,minimum size=0.3cm] at (-1.5, 0) {$SD_1$};
    \draw[->] (-2.1, -1.3)--(-2.1, 1.3);
    \node[align=center,minimum size=0.3cm] at (-2.9, 0) {$M_3(SD_1)$};
    \draw[->] (-1.9, 1.3)--(-1.9, -1.3);
    
    \node[align=center,minimum size=0.3cm] at (0.3, 0.3) {$SD_1$};
    \draw[->] (-1.45, 1.55)--(1.55, -1.45);
    \node[align=center,minimum size=0.3cm] at (-.5, -.5) {$M_4(SD_1)$};
    \draw[->] (1.45, -1.55)--(-1.55, 1.45);
    
    \draw[draw=blue!80!white, dashed, rounded corners] (-6.3, 4.4) rectangle (4.8, 3.1);
    
    \node[align=center,minimum size=0.3cm] at (-3.1, 4) {Sharing level $\lambda_1 = \vert SD_1\vert / \vert D_1\vert$};
    \draw[draw=green!80!white, rounded corners] (-6.2, 3.8) rectangle (0, 4.25);
    
    \node[align=center,minimum size=0.3cm] at (-3.1, 3.5) { $\left[M_1(SD_1)M_2(SD_1)M_3(SD_1)M_4(SD_1)\right]$};
    \draw[draw=green!80!white, rounded corners] (-6.2, 3.25) rectangle (0, 3.75);
    
    \draw[fill=orange!80!white, rounded corners] (1.35, 3.2) rectangle (4.7, 3.8);
    \node[align=center,minimum size=0.3cm] at (3.05, 3.5) {$[c_1^2,c_1^3,c_1^4],\Sigma_{j=2}^4c_1^j=1$};
    
    \draw[blue,->] (0, 3.5)--(1.35, 3.5);
    \node[align=center,minimum size=0.3cm] at (0.7, 3.5) {majority \\ voting};
    
    \draw[blue,->] (-2, 2.7)--(-2, 3.1);
    
    \node[align=center,minimum size=0.3cm] at (-3.6, 2.8) {DPGAN, $SD_1$};
    \dpgan{(-3.6, 2)};
    \node[align=center,minimum size=0.3cm] at (3.6, 2.8) {DPGAN, $SD_2$};
    \dpgan{(3.6, 2)};
    \node[align=center,minimum size=0.3cm] at (-3.6, -1.2) {DPGAN, $SD_3$};
    \dpgan{(-3.6,-2)};
    \node[align=center,minimum size=0.3cm] at (3.6, -1.2) {DPGAN, $SD_4$};
    \dpgan{(3.6,-2)};


    \draw[draw=blue!80!white,dashed, ultra thick] (5,-3) -- (5,5);
    
    \node[text=orange!100!white] at (0,4.7) {1st Stage: Local credibility and tokens initialisation};
    
    \node[text=orange!100!white] at (9.5,4.7) {2nd Stage: Local credibility and tokens update};

    
    \draw[draw=cyan] (-2+10, 2) circle (0.7);
    \draw[draw=cyan] (2+10, 2) circle (0.7);
    \draw[draw=cyan] (-2+10, -2) circle (0.7);
    \draw[draw=cyan] (2+10, -2) circle (0.7);
    
    \node[align=center,minimum size=0.3cm] at (-2+10, 2.05) {Party 1 \\ \small $\Delta w_1$,$M'_1$};
    \node[align=center,minimum size=0.3cm] at (2+10, 2.05) {Party 2 \\ \small $\Delta w_2$,$M'_2$};
    \node[align=center,minimum size=0.3cm] at (-2+10, -1.95) {Party 3 \\ \small $\Delta w_3$,$M'_3$};
    \node[align=center,minimum size=0.3cm] at (2+10, -1.95) {Party 4 \\ \small $\Delta w_4$,$M'_4$};

    \node[align=center,minimum size=0.3cm] at (0+10, 2.3) {$(\Delta w_1^2)^S$};
    \draw[->] (-1.3+10, 2.1)--(1.3+10, 2.1);
    \node[align=center,minimum size=0.3cm] at (0+10, 1.7) {$(\Delta w_2^1)^S$};
    \draw[->] (1.3+10, 1.9)--(-1.3+10, 1.9);
    
    \node[align=center,minimum size=0.3cm] at (-1.3+10, 0) {$(\Delta w_1^3)^S$};
    \draw[->] (-2.1+10, -1.3)--(-2.1+10, 1.3);
    \node[align=center,minimum size=0.3cm] at (-2.7+10, 0) {$(\Delta w_3^1)^S$};
    \draw[->] (-1.9+10, 1.3)--(-1.9+10, -1.3);
    
    \node[align=center,minimum size=0.3cm] at (0.5+10, 0.3) {$(\Delta w_1^4)^S$};
    \draw[->] (-1.45+10, 1.55)--(1.55+10, -1.45);
    \node[align=center,minimum size=0.3cm] at (-.3+10, -.6) {$(\Delta w_4^1)^S$};
    \draw[->] (1.45+10, -1.55)--(-1.55+10, 1.45);
    
    \draw[draw=blue!80!white, dashed, rounded corners] (-4.7+10, 4.4) rectangle (0.7+10, 3.1);
    
    \draw[draw=green!80!white, rounded corners] (-4.5+10, 3.78) rectangle (0.5+10, 4.25);
    \node[align=center,minimum size=0.3cm] at (-2+10, 4) {$w'_1 = w_1 + \Delta w_1 + \Sigma_{j=2}^4(\Delta w_j^1)^S$};
    
    \draw[draw=green!80!white, rounded corners] (-4.5+10, 3.25) rectangle (0.5+10, 3.72);
    \node[align=center,minimum size=0.3cm] at (-2+10, 3.5) {\small $[{c_1^2}',{c_1^3}',{c_1^4}'],\Sigma_{j=2}^4{c_1^j}'=1$};
    
    \draw[blue,->] (-2+10, 2.7)--(-2+10, 3.1);
\end{tikzpicture}}
\caption{Two-stage Realization of FDPDDL. $SD_1$: DPGAN samples randomly chosen by Party 1 from the pool of local DPGAN samples generated offline, $M_1$: standalone model of Party 1; $(\Delta w_2^1)^S$: selected gradients of Party 2 sent to Party 1 $(d_2^1=\min(c_1^2\times d_1,\lambda_2\times\vert\Delta w_2\vert)$ gradients are selected from $\Delta w_2$ and grouped into set $S$, where $c_1^2\times d_1$ is the download request from Party 1), $M'_2$: local model of Party 2 at current communication round.}
\label{fig:FDPDDL_flow}
\vspace{-3mm}
\end{figure*}
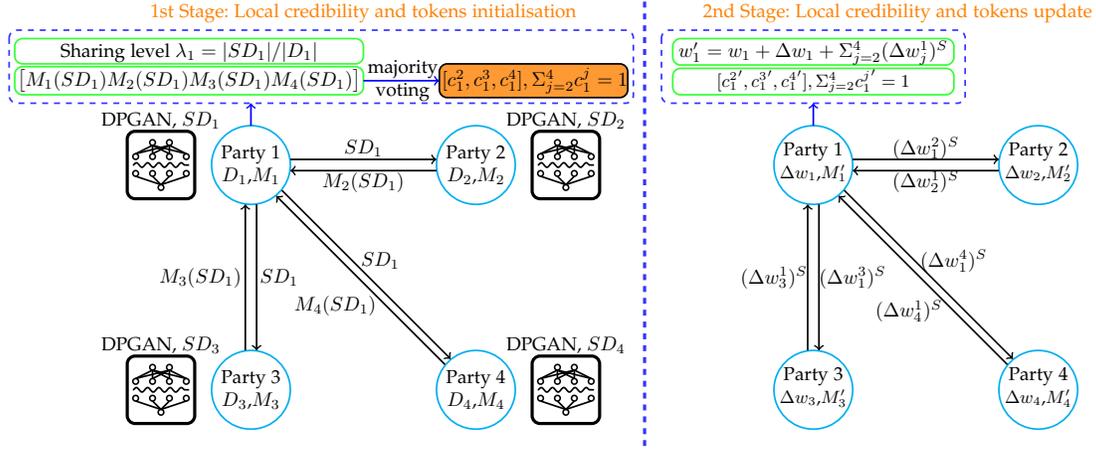

\subsection{Local credibility and tokens initialisation}\label{sec:benchmarking}
\begin{algorithm}[ht]
\caption{Local credibility and tokens initialisation}\label{Algorithm:benchmarking}
\small
\begin{algorithmic}
  \State \textbf{Input: number of participating parties $n$, C=\{1,\ldots,n\}}
  \State \textbf{Output: local credibility and tokens of all parties}
  \State 1: \textbf{Pre-train aprior model}: Each party $i$ trains standalone model $M_i$ based on its local training data.
   \State 2: \textbf{Artificial samples generation}: Party $i$ releases $u_i=\lambda_i*|D_i|$ artificial samples generated by DPGAN to any party $j$.
   \State 3: \textbf{Local credibility initialisation}: Party $j$ labels the received artificial samples by its standalone model $M_j$, then returns the predicted labels back to party $i$. Meanwhile, party $i$ also predicts labels for its own DPGAN samples using $M_i$. Party $i$ then applies majority voting to all the predicted labels, and initialises the local credibility of party $j$ as $c_i^j=\frac{m_j}{u_i}$, where $m_j$ is the number of matches between majority labels and party $j$'s predicted labels, and $u_i$ is the number of DPGAN samples generated by party $i$. 
   \State 4: \textbf{Local credibility normalisation}: 
      $c_i^j=\frac{c_i^j}{{\textstyle\sum}_{j \in C \setminus i} c_i^j}$
      \If{$c_i^j<c_{th}$}
      \State party $i$ reports party $j$ as "non-credible"
     \EndIf
    \State 5: \textbf{Credible party set}: If majority party report party $j$ as "non-credible", Blockchain removes party $j$ from the credible party set $C$ and all parties rerun step 4 again.
 \State 6: \textbf{Tokens initialisation to download gradients}: $p_i=\lambda_i*|\vec{w}_i|*(n-1)$.
\end{algorithmic}
\end{algorithm}

As stated in Algorithm~\ref{Algorithm:benchmarking}, to initialise local credibility and tokens, each participant first trains a DPGAN based on its local training data to generate artificial samples with differential privacy guarantee. These artificial samples are generated in a way that \textbf{does not} disclose the true sensitive image instances, as well as the true distribution of data. Rather, they only provide a few implicit density estimation within a tolerable privacy budget used in DPGAN~\cite{zhang2018differentially}. Each participant then publishes individually generated samples with size proportional to individual sharing level without publishing any labels. After receiving DPGAN samples from one participant, all the other participants run their pre-trained standalone models on these received artificial samples and send the predicted labels back to the sender for local credibility initialisation. 
Below, we detail the main tasks in Algorithm
\ref{Algorithm:benchmarking}: sharing level and digital tokens initialisation, and local credibility initialisation according to the number of released artificial samples and relative contribution of each party. 

\textbf{Sharing level and Tokens Initialisation:} Based on the number of artificial samples $u_i$ that party $i$ publishes at the beginning, sharing level is autonomously determined that it is comfortable with, which can be quantified as $\lambda_i=u_i/|D_i|$, where $D_i$ is the local training data of party $i$. The more private party would prefer to release less samples, while the less private party is comfortable with releasing more samples. Similarly, during the update stage, more private party would prefer to release less gradients. Tokens of party $i$ are initialised as
$p_i=\lambda_i*|\vec{w}_i|*(n-1)$, 
where $\lambda_i$ is the sharing level of party $i$, $|\vec{w}_i|$ is the number of model parameters, and $n$ is the number of parties. The gained tokens will be used to download gradients in the update stage.

\textbf{Local Credibility Initialisation.} For local credibility initialisation, each party compares the majority voting of all the combined labels with an individual party's predicted labels to evaluate the effect of this party. It relies on the fact that the majority voting of all the combined labels reflects the outcome of the majority party, while the predicted labels of party $j$ only reflects the outcome of party $j$. For example, in the case of party $i$ initialising local credibilities for other parties, party $i$ broadcasts its artificial samples generated by DPGAN to other parties, who label these samples using their pre-trained standalone models, then send the corresponding predicted labels back to party $i$. Meanwhile, party $i$ also labels its own artificial samples using its pre-trained standalone model, then combines all the predicted labels of all parties as a label matrix with total $n$ columns with each column corresponding to one party's predicted labels. From this label matrix, party $i$ can initialise the local credibility of party $j$ as
$c_i^j=\frac{m_j}{u_i}$,
where $m_j$ is the number of matches between the majority labels and party $j$'s predicted labels, and $u_i$ is the number of DPGAN samples released by party $i$. Afterwards, party $i$ normalises $c_i^j$ within [0,1]. If the majority party report that the local credibility of one party is lower than the threshold $c_{th}$, implying a ``non-credible'' party, it will be banned from the local credibility lists of all parties. Here, $c_{th}$ should be agreed by the majority party.

In addition, Algorithm~\ref{Algorithm:benchmarking} can automatically take care of the scenario where an honest participant publishes some gradients, while all the other honest participants assign very low credibility. In this case, the data distribution of the publisher is completely different from that of the other participants, hence it is still reasonable to reduce the credibility of the publisher, because other participants are anyway unlikely to gain much from the updates released by the publisher.

\begin{figure}[!htp]
\centering
        \begin{subfigure}[ht]{0.23\textwidth}
                \includegraphics[width=3.8cm,height=2cm]{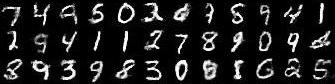}\label{fig:MNIST_gan_sample}
        \end{subfigure}
        \begin{subfigure}[ht]{0.23\textwidth}
                \includegraphics[width=3.8cm,height=2cm]{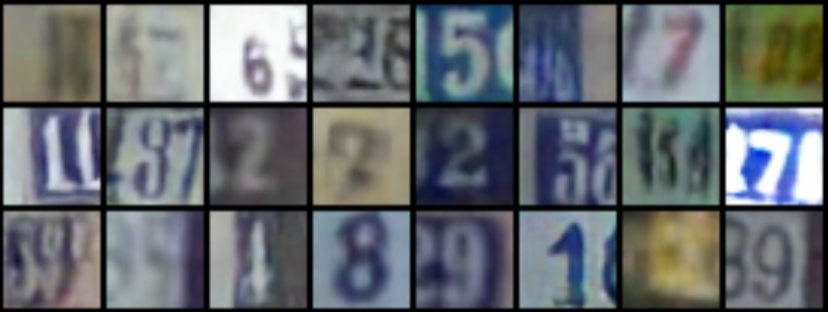}\label{fig:SVHN_gan_sample}
        \end{subfigure}
\caption{Generated samples by DPGAN with $\epsilon=4, \delta=10^{-5}$ for MNIST and $\epsilon=4, \delta=10^{-6}$ for SVHN. Each party trains a DPGAN on 60,000 augmented MNIST examples, and 100,000 augmented SVHN examples respectively.}
\label{fig:dpgan_samples}
\vspace{-3mm}
\end{figure}

\textbf{Differentially Private GAN (DPGAN).}
During the initialisation stage, we use \emph{Differentially Private Generative Adversarial Network} (DPGAN) to generate differentially private artificial samples to mutually benchmark the local credibility of each party and generate initial tokens. Each party individually trains a \emph{Differentially Private GAN} (DPGAN) by using GANobfuscator which adds tailored noise to gradients during the training procedure of GAN~\cite{xu2019ganobfuscator}. The main idea lies in the post-processing property of differential privacy, differentially private discriminator combined with the computation of generator will produce differentially private generator. To counter the stability and scalability issues of training DPGAN, we apply adaptive pruning, which significantly improve both training stability and utility~\cite{xu2019ganobfuscator}. DPGAN can generate infinite number of samples for the intended analysis, while rigorously guaranteeing $(\epsilon,\delta)$-differential privacy of training data. Without loss of generality, we exemplify DPGAN in the context of the improved WGAN framework~\cite{arjovsky2017wasserstein}. As demonstrated by the most recent work~\cite{zhang2018differentially,xu2019ganobfuscator}, DPGAN is able to synthesize data with inception scores fairly close to both the real data and the samples generated by the non-private GANs. As evidenced by Fig.~\ref{fig:dpgan_samples}, although the generated artificial samples are not real training samples, the digits clearly vary either in shape, colour or surroundings, they can still keep the general characteristics of the class to ensure utility. Due to limited data size of each party, we let each party apply data augmentation to expand local data size to 100 times, which helps DPGAN to generate more reliable samples within a moderate privacy budget for local credibility initialisation. In particular, we augment image datasets with rotation range of 1 and width shift range and height shift range of 0.01. For text datasets, we repeat each record for 100 times. Samples generated by DPGAN with $\epsilon=4$ and $\delta=10^{-5}$ for MNIST and $\epsilon=4, \delta=10^{-6}$ for SVHN are illustrated in Fig.~\ref{fig:dpgan_samples}. Each party individually trains a DPGAN on 60,000 augmented MNIST examples, and 100,000 augmented SVHN examples respectively. Note that each party can generate massive DPGAN samples offline without affecting collaboration.
 
\subsection{Local credibility and tokens update}
\label{sec:credit_update}
For local credibility and tokens update, each part $i$ takes 20\% local training data as the validation data and uses leave-one-out strategy to evaluate the local credibility of party $j$ based on the usefulness of party $j$'s gradients in each round of training process. Specifically, party $i$ evaluates the change of validation accuracy by removing party $j$'s gradients from the updated model parameter $\vec{w}_i'$ that combines all parties' gradients, \ie using the combined gradients with and without party $j$'s gradients to evaluate the validation accuracy of party $i$, which yield $acc$ and $acc_j$ respectively, the difference between $acc$ and $acc_j$ reflects how party $j$ affects validation accuracy. Party $i$ computes the local credibility $c_i^j$ of party $j$ at the current round by passing an "accuracy factor" $x=\frac{acc}{acc+acc_j}$ through a sigmoid function $f$ as in Eq.~\eqref{eq:credibility}.
\begin{equation}\label{eq:credibility}
	c_i^j=f(x)=\frac{1}{1+exp(-15*(x-0.5))}
\end{equation}
The incentive can be explicitly explained as follows. As $x$ stands for the accuracy ratio between the validation accuracy using the combined gradients of all parties and the validation accuracy using the combined gradients without party $j$'s gradients, hence it can be further expressed as:
 \begin{equation}\label{eq:factor}
x=\frac{acc}{acc+acc_j}=\frac{acc}{2*acc+\Delta}
\end{equation}
where $acc_j=acc+\Delta$, $\Delta$ indicates the impact of removing party $j$, the more positive the value of $\Delta$, the better the validation accuracy after removing party $j$, hence the lower the contribution of party $j$. To be more specific, if party $j$ has no impact, $\Delta=0, x=0.5, c_i^j=0.5$; if party $j$ contributes negatively, $acc_j>acc$, then $\Delta>0, x<0.5, c_i^j<0.5$; if party $j$ contributes positively, $acc_j<acc$, then $\Delta<0, x>0.5, c_i^j>0.5$. Each party $i$ keeps updating its local credibility list based on the contributions of all the other parties in each round and integrates their historical local credibilities by averaging over the local credibility of current round and previous round. In the follow-up rounds, the number of gradients to be downloaded will be dependent on the sharing level, local credibility and download budget.

For tokens update, one token is consumed/rewarded for each download/upload of gradients. In the subsequent rounds, party $i$ is more likely to download gradients from more credible parties, while download less, even ignore those published by less credible parties. If the credibility of one party falls below a threshold $c_{th}$, it can be even banned from the local credibility list of party $i$. 

\begin{algorithm}[ht]
\caption{Local credibility and tokens update}\label{Algorithm:local_credibility_update}
\small
\begin{algorithmic} 
\State \textbf{Input: $C$, $c_i^j$, $p_i$, $p_j$, $d_i$, $\lambda_j$, $\Delta \vec{w}_j$, $\vec{w}_i$, $V_i$}
\State \textbf{Output: updated parameters $\vec{w}_i'$, credibility ${c_i^j}'$ and tokens $p_j'$, $p_i'$}
\State In each round, suppose party $i$ aims to download total $d_i$ gradients from all parties in $C$, while party $j \in C \setminus i$ can at most upload $\lambda_j \times |\Delta \vec{w}_j|$ gradients. Party $i$ updates its local credibility list, model parameters and tokens based on the gradients of party $j \in C \setminus i$ as follows:
      
\If{$d_i<p_i$}
\For{$j \in C \setminus i$}
\State $d_j^i=min(c_i^j*d_i, \lambda_j*|\Delta \vec{w}_j|)$
\State $p_j'=p_j+d_j^i$, $p_i'=p_i-d_j^i$
\State $d_j^i$ gradients of $\Delta \vec{w}_j$ are grouped into set $S$, which are selected according to the ``largest values" criterion: sort gradients in $\Delta \vec{w}_j$, and upload $d_j^i$ of them, starting from the largest.

\State \textbf{Parameter update}:  $\vec{w}_i'=\vec{w}_i+\Delta {\vec{w}_i}+{\textstyle\sum}_{j \in C \setminus i} \Delta (\vec{w}_j^i)^S$, ${\vec{w}_i^j}'=\vec{w}_i'-\Delta (\vec{w}_j^i)^S$, where $\vec{w}_i$ is party $i$'s local parameters of previous communication round.

\State $acc \leftarrow (\vec{w}_i', V_i)$, $acc_j \leftarrow ({\vec{w}_i^j}', V_i)$
\State $x=\frac{acc}{acc+acc_j}$
\State ${c_i^j}'=\frac{c_i^j+f(x)}{2}$, where $f$ refers to the sigmoid credibility mapping function in Eq.~\eqref{eq:credibility}.
\EndFor
\State \textbf{credibility normalisation}: 
      ${c_i^j}'=\frac{{c_i^j}'}{{\textstyle\sum}_{j \in C \setminus i} {c_i^j}'}$
    \If{${c_i^j}'<c_{th}$}
	\State party $i$ reports party $j$ as "non-credible"
    \EndIf
\State \textbf{Credible party set}: If majority party report party $j$ as "non-credible", Blockchain removes party $j$ from credible party set $C$ and all parties remove party $j$'s model updates $\Delta (\vec{w}_j^i)^S$ from their updated $\vec{w}_i'$ and rerun credibility normalisation.
\EndIf
\end{algorithmic}
\end{algorithm}

The detailed local credibility and tokens update procedure is elaborated in Algorithm~\ref{Algorithm:local_credibility_update}. Note that for any party $i$, the received gradients ${\textstyle\sum}_{j \in C \setminus i} d_j^i$ could be different from the download budget $d_i$, as party $j$ at most can provide $\lambda_j*|\Delta \vec{w}_j|$ gradients, while party $i$ plans to download $c_i^j*d_i$ gradients from party $j$. To fill in the gap between $d_i$ and ${\textstyle\sum}_{j \in C \setminus i} d_j^i$, we design a supplement mechanism which can be referred to the supplementary material.

\textbf{Differentially Private SGD (DPSGD):}
To facilitate collaborative privacy-preserving deep learning, we use DPSGD~\cite{abadi2016deep,mcmahan2018learning} to enable information exchange in a differentially private manner. DPSGD consists of two parts: sanitizer and moments accountant. Sanitizer performs two operations: (1) limit the sensitivity of each individual example by clipping the norm of its gradient; and (2) add noise to the gradient of a lot (several mini-batches) before updating network parameters. Moments accountant keeps track of a bound on the moments of the privacy loss random variable to compute the spent privacy over the course of training. Different from DPSGD used for the whole database in the centralised framework~\cite{abadi2016deep}, decentralised framework enables each party to individually train local model, and we are concerned with the privacy leakage from the local model before publication in each round. To limit the sensitivity of updates, we follow the DPSGD algorithm~\cite{abadi2016deep} to clip the gradient of each example such that the $L_2$ norm is bounded by the chosen gradient norm upper bound. Model training that satisfies differential privacy with respect to example-adjacent datasets satisfies the intuitive notion of privacy: the presence or absence of any specific example in the training data has an imperceptible impact on the parameters of the learned model~\cite{beaulieu2018privacy}. It follows that an adversary inspecting the trained model cannot infer whether any specific example was used in the training, irrespective of what auxiliary information they may have.

We choose $\sigma=\sqrt{2*ln(1.25/\delta)}/\epsilon$ for DPSGD, where $\epsilon \leq 1$, by the standard arguments~\cite{dwork2014algorithmic}, each step is $(\epsilon,\delta)$-DP with respect to each lot.
Since each lot is randomly sampled with replacement, the privacy amplification theorem~\cite{beimel2010bounds} implies each step is $(O(q\epsilon),q\delta)$-DP w.r.t the full database, where $q = L/N$ is the sample ratio. 
Compared with strong composition theorem~\cite{dwork2010boosting}, moments accountant delivers tighter bound in two ways~\cite{abadi2016deep}: it saves a $\sqrt{log(1/\delta)}$ factor in the $\epsilon$ part and a $Tq$ factor in the $\delta$ part. For appropriately chosen noise scale and clipping threshold, DPSGD is $(O(q\epsilon \sqrt{T},\delta))$-differentially private. Here, $T$ is the total number of iterations over the training data. Because of this tighter bound on privacy spending, DPSGD can iterate over the training data sufficient number of times before exhausting a moderate privacy budget. This explains why DPSGD is able to train deep models that offer good model utility.

We remark that one attractive consequence of applying DPGAN and DPSGD is that integrating DP into training generalizes well~\cite{bassily2016algorithmic}. Like normal GAN and normal SGD, DPGAN and DPSGD need to iterate over the training data and apply gradient computation multiple times. However, each access to the training data causes information leakage of the training data and thus incurs privacy loss from the overall privacy budget $\epsilon$. 
To apply DPGAN and DPSGD to the distributed/decentralized settings, we follow recent work~\cite{zhang2016dynamic,huang2019dp,kim2019efficient,lyu2019fog} to conduct local gradient computation and calculate privacy on a per party-basis, where each party individually applies moments accountant~\cite{abadi2016deep} to keep track of the spent privacy budget. Each party repeats the local training process until the allocated privacy budget is used up. In particular, for local training process of DPGAN and DPSGD over local dataset of each party, we allocate a privacy budget of $(4,10^{-5})$-DP and $(2,10^{-5})$-DP respectively (with the exception of SVHN where $\delta= 10^{-6}$). As per composition property of DP in Theorem~\ref{Theorem_dpcom}, it results in a total $(6,2*10^{-5})$-DP for MNIST, Adult and Hospital, and $(6,2*10^{-6})$-DP for SVHN for each party.

\section{Performance Evaluation}
\label{sec:Performance}

\subsection{Datasets}
\emph{MNIST\footnote{\url{http://yann.lecun.com/exdb/mnist/}}}. This dataset is for handwritten digit recognition, which consists of 60,000 training examples and 10,000 test examples. Each example is a 32x32 gray-level image, with digits locating at the center of the image. 

\emph{SVHN\footnote{\url{http://ufldl.stanford.edu/housenumbers/}}}. This dataset is obtained from Google's street view images, containing over 600,000 examples, from which we use 100,000 for training and 10,000 for testing. Each example is a 32x32 centered image with RGB channels.
SVHN is more challenging than MNIST as most images are noisy and contain distractors at the sides.
The classification objective for both MNIST and SVHN is to classify the input image as one of 10 possible digits within [``0''-``9''].

\emph{Adult\footnote{\url{http://archive.ics.uci.edu/ml/datasets/Adult}}}. 
The Adult Census dataset includes 48,843 records with 14 sensitive attributes, including age, race, education level, marital status, and occupation, etc. This dataset is commonly used to predict whether an individual makes over 50K dollars in a year (binary). There are 48,842 records in total, with 24\% (11,687) records over 50K and 76\% (37,155) under 50K. We manually balance the dataset to 11,687 records over 50K and 11687 records under 50K by random sampling, resulting in 23,374 records. We allocate 80\% records as training set and 20\% as test set.

\emph{Hospital}. The Diabetic Hospital dataset contains data on diabetic patients from 130 US hospitals and integrated delivery networks. We directly derived the dataset from~\cite{koh2017understanding} which balances the training set to 10k positives and 10k negatives. The record of each patient is represented by 127 features, such as demographic (\eg gender, race, age), administrative (\eg length of stay) and medical (\eg test results).
The task is to predict whether a patient would be readmitted to hospital within 30 days (binary). 

\subsection{SGD Frameworks}
To show the effectiveness of our proposed FDPDDL, we compare with three baselines as outlined in~\cite{shokri2015privacy}. SGD is adopted in all frameworks. 

\textbf{Centralised framework} assumes all the local training data are pooled into a trusted server to train a global model on the combined data using standard SGD.

\textbf{Standalone framework} enables participants to train standalone models on their local training data without any collaboration. When training alone, each participant is susceptible to falling into local optima. 

\textbf{Distributed framework} allows participants to train independently and concurrently, and to choose a fraction of parameters to upload per round. Distributed framework using selective SGD (DSSGD) can achieve equivalent or even higher performance than the centralised framework because updating a small fraction of parameters acts as a regularisation technique, which prevents the neural network from "memorizing" training data, hence avoiding overfitting~\cite{shokri2015privacy}. Therefore, we also use DSSGD in distributed framework. As DSSGD with round robin parameter exchange protocol results in the highest accuracy in~\cite{shokri2015privacy} and facilitates fairness calculation, we follow the round robin protocol for DSSGD, where participants run SSGD sequentially: a party downloads a fraction of the most up-to-date parameters from the server, runs local training, and uploads selected gradients; the next party follows in the fixed order~\cite{shokri2015privacy}.
Gradients are selected and uploaded according to the ``largest values'' criterion which is consistent throughout the entire learning process in DSSGD. 

\subsection{Communication Protocol}
Asynchronous protocol may lead to concurrency issues, also known as the staleness effect~\cite{zhang2015staleness}, which is due to the speed inconsistency between different parties. For those slow parties, downloaded parameters may lose usefulness if other parties perform parameter updates much more frequently. This staleness effect can slow down convergence or even destroy learning.

In contrast, synchronous SGD typically works better than the asynchronous SGD, as demonstrated in~\cite{chen2016revisiting}, synchronous training achieves around 0.5\% to 0.9\% higher accuracy, needs fewer epochs to converge, and scales better. Therefore, we use synchronous parameter exchange protocol in all our experiments. However, we observed that both federated learning~\cite{mcmahan2017communication,bonawitz2017practical} and distributed learning with DSSGD~\cite{shokri2015privacy} suffer from certain non-convergence and accuracy degradation problems working with this synchronous protocol. That partly explains why DSSGD adopts asynchronous, round robin or random order protocols, rather than synchronous protocol~\cite{shokri2015privacy}. We also observed that our FDPDDL framework is less sensitive to various hyper-parameter settings, \eg it \emph{does not} suffer from non-convergence problem even using the same hyper-parameter setting as in DSSGD. We hypothesize that the downloading strategy based on accumulated credibility contributes to model convergence -- a by-product of our framework.

\subsection{Experimental Setup}
\label{sec:Setup}
For implementation on image datasets including MNIST and SVHN, we use \emph{multi-layer perceptron} (MLP) and \emph{convolutional neural network} (CNN) architectures as in~\cite{shokri2015privacy}. The detailed architecture description is deferred to the supplementary file. For text datasets including Adult and Hospital, we use MLP with a single hidden layer with 128 units. To reduce the impact of random initialisation and counter non-convergence, each party initialises its local model with the same parameter $\vec{w}_0$, then runs training on its local data to update local model parameter $\vec{w}_i$. This contributes to a fair and consistent local credibility initialisation. For local model training, we follow the preliminary study in~\cite{abadi2016deep} to choose the lot size as $\sqrt{N}$, where $N$ is the total number of local training examples including the augmented examples in our case, and set the initial learning rate as $0.1$ with decay $10^{-7}$. During the training of DPGAN and DPSGD, we dynamically adjust the clipping bounds to achieve faster convergence and better utility~\cite{xu2019ganobfuscator}. 
To boost fairness and enable local models to move towards their respective model minima, we let each party individually train 10 local epochs in advance before collaborative learning starts. For all the experiments, we empirically set the local credibility threshold as $c_{th}=\frac{1}{n}*\frac{2}{3}$ via grid search, where $n$ is the number of parties. 
 
For applicability, we mainly investigate three realistic settings as follows, among which, the first two settings belong to the balanced partition, while the last setting belongs to the unbalanced partition. In particular, for the balanced partition of image datasets, we randomly sample $1\%$ of the entire examples as the local training data for each party, \ie $600$ examples for MNIST and $1000$ examples for SVHN; for the balanced partition of text datasets, we randomly sample 370 examples as the local training data of each party for Adult dataset, and 400 examples as the local training data of each party for Hospital dataset. 

\begin{itemize}
\item \textbf{Same sharing level $\lambda_i$, same data size $|D_i|$:} we set the sharing level of each party as $0.1$, where each party releases $10\%$ artificial samples during the initialisation stage, and $10\%$ gradients during the update stage;

\item \textbf{Different sharing level $\lambda_i$, same data size $|D_i|$:} we randomly assign sharing levels from $[0.1,0.5]$ to each party, each party releases artificial samples and gradients as per their individual sharing levels; 

\item \textbf{Different data size $|D_i|$, same sharing level $\lambda_i$:} the difference of this setting from the previous two settings lies in that different parties are allocated with different number of examples, \ie imbalanced data size. For example, for MNIST, we randomly partition a total of \{2400, 9000, 18000, 30000\} examples among \{4, 15, 30, 50\} parties respectively. Similarly, for SVHN, a total of \{4000, 15000, 30000, 50000\} examples are randomly partitioned among \{4, 15, 30, 50\} parties respectively. The sharing level of each party is set equally to $0.1$.
\end{itemize}

\subsection{Quantification of Fairness}
In collaborative learning, collaborative fairness should be quantified from the view of the whole system. In this work, we quantify collaborative fairness through the correlation coefficient between party contributions (\ie standalone model accuracies which characterize the learning capability of different parties on their own data, and sharing levels, which characterize the sharing willingness of different parties) and party rewards (\ie final model accuracies of different parties). 

Specifically, we take party contributions as the X-axis, which represents the contributions of different parties from the system view. In particular, in Setting 2, we characterize different parties' contributions by their sharing levels and standalone model accuracies, as the party who is less private and has local data with better generalization empirically contributes more. In Setting 1 and Setting 3, we characterize different parties' contributions by their standalone model accuracies, as the party who has local data with better generalization empirically contributes more. Specifically, in Setting 3, the party with more local data typically yields higher standalone model accuracy in IID scenarios. In summary, the X-axis can be expressed by Eq.~\ref{eq:x_axis}, where $\lambda_j$ and $sacc_j$ denote the sharing level and standalone model accuracy of party $j$ respectively: 
\begin{equation}\label{eq:x_axis}
\vec{x}=
\resizebox{0.93\hsize}{!}{
$\Big\{\begin{array}{cl}
\{\frac{\lambda_1}{\textstyle\sum \lambda_j},\cdots,\frac{\lambda_n}{\textstyle\sum \lambda_j}\}+\{\frac{sacc_1}{\textstyle\sum sacc_j},\cdots,\frac{sacc_n}{\textstyle\sum sacc_j}\},  & \mbox{Setting 2} \\
\{sacc_1,\cdots,sacc_n\}, & \mbox{Setting 1\&3} 
\end{array}
$}
\end{equation}

Similarly, we take party rewards (\ie final model accuracies of different parties) as the Y-axis, $\vec{y}=\{acc_1,\cdots,acc_n\}$, where $acc_j$ denotes the final model accuracy of party $j$.

As the Y-axis measures local model performance of different parties after collaboration, it is expected to be positively correlated with the X-axis to deliver good fairness. Hence, we formally quantify collaborative fairness in Eq.~\ref{eq:fairness}:
\begin{equation}\label{eq:fairness}
r_{xy}=\frac{{\textstyle\sum}_{i=1}^n (x_i-\bar{x})(y_i-\bar{y})}{(n-1)s_xs_y}
\end{equation}
where $\bar{x}$ and $\bar{y}$ are the sample means of $\vec{x}$ and $\vec{y}$, $s_x$ and $s_y$ are the corrected standard deviations. The range of fairness is within [-1,1], with higher values implying good fairness. Conversely, negative coefficient implies poor fairness. 

\subsection{Experimental Results}
\begin{table}[ht]
\caption{Fairness test of distributed framework and FDPDDL on MNIST, with different party numbers (P-$k$) and different settings.}
\label{tbl:MNIST_fairness}
\centering
\resizebox{\linewidth}{!}{
\begin{tabular}{p{0.25cm}cccccccc}
\multirow{2}{*}{} & \multicolumn{4}{c}{Different $\lambda_i$, same $|D_i|$} & \multicolumn{4}{c}{Different $|D_i|$, same $\lambda_i$}
\tabularnewline
\cmidrule(lr){2-5}\cmidrule(lr){6-9}
 & \multicolumn{2}{c}{Distributed} & \multicolumn{2}{c}{FDPDDL} & \multicolumn{2}{c}{Distributed} & \multicolumn{2}{c}{FDPDDL}
\tabularnewline
\cmidrule(lr){2-3}\cmidrule(lr){4-5}\cmidrule(lr){6-7}\cmidrule(lr){8-9}
 & CNN & MLP & CNN & MLP & CNN & MLP & CNN & MLP
\tabularnewline
\midrule
\textit{P4}  &-0.68 &0.30 &0.92 &0.96 &-0.97 &0.28 &0.95 &0.98
\tabularnewline
\textit{P15} &0.20 &-0.15 &0.90 &0.92  &0.03 &-0.07  &0.91 &0.90
\tabularnewline
\textit{P30} &-0.02 &0.02 &0.87 &0.85  &0.04 &0.13 &0.84 &0.78
\tabularnewline
\textit{P50} &-0.16 &-0.05 &0.78 &0.76  &0.14 &0.07 &0.75 &0.71
\tabularnewline
\bottomrule
\end{tabular}
}
\vspace{-3mm}
\end{table}

\begin{table}[ht]
\caption{Fairness test on SVHN.}
\label{tbl:SVHN_fairness}
\centering
\resizebox{\linewidth}{!}{
\begin{tabular}{p{0.25cm}cccccccc}
\multirow{2}{*}{} & \multicolumn{4}{c}{Different $\lambda_i$, same $|D_i|$} & \multicolumn{4}{c}{Different $|D_i|$, same $\lambda_i$}
\tabularnewline
\cmidrule(lr){2-5}\cmidrule(lr){6-9}
 & \multicolumn{2}{c}{Distributed} & \multicolumn{2}{c}{FDPDDL} & \multicolumn{2}{c}{Distributed} & \multicolumn{2}{c}{FDPDDL}
\tabularnewline
\cmidrule(lr){2-3}\cmidrule(lr){4-5}\cmidrule(lr){6-7}\cmidrule(lr){8-9}
 & CNN & MLP & CNN & MLP & CNN & MLP & CNN & MLP
\tabularnewline
\midrule
\textit{P4}  &0.27 &0.26 &0.89 &0.85 &0.38 &0.20 &0.98 &0.97
\tabularnewline
\textit{P15} &0.16 &0.19 &0.83 &0.79  &-0.13 &0.36  &0.90 &0.89
\tabularnewline
\textit{P30} &-0.14 &0.12 &0.75 &0.69  &0.04 &-0.27 &0.85 &0.84
\tabularnewline
\textit{P50} &-0.25 &-0.37 &0.72 &0.66  &-0.23 &0.15 &0.77 &0.73
\tabularnewline
\bottomrule
\end{tabular}
}
\vspace{-3mm}
\end{table}

\begin{table}[ht]
\caption{Fairness test on Adult and Hospital.}
\label{tbl:AdultHospital_fairness}
\centering
\resizebox{\linewidth}{!}{
\begin{tabular}{p{0.25cm}cccccccc}
\multirow{2}{*}{} & \multicolumn{4}{c}{Different $\lambda_i$, same $|D_i|$} & \multicolumn{4}{c}{Different $|D_i|$, same $\lambda_i$}
\tabularnewline
\cmidrule(lr){2-5}\cmidrule(lr){6-9}
 & \multicolumn{2}{c}{Distributed} & \multicolumn{2}{c}{FDPDDL} & \multicolumn{2}{c}{Distributed} & \multicolumn{2}{c}{FDPDDL}
\tabularnewline
\cmidrule(lr){2-3}\cmidrule(lr){4-5}\cmidrule(lr){6-7}\cmidrule(lr){8-9}
 &Adult &Hosp &Adult &Hosp &Adult &Hosp &Adult &Hosp
\tabularnewline
\midrule
\textit{P4} &0.13 &0.15 &0.97 &0.94 &0.15 &0.18 &0.99 &0.95
\tabularnewline
\textit{P15} &0.02 &0.07 &0.90 &0.85 &0.07 &0.10 &0.92 &0.88
\tabularnewline
\textit{P30} &-0.08 &-0.12 &0.75 &0.71 &-0.02 &0.03 &0.77 &0.74
\tabularnewline
\textit{P50} &-0.12 &-0.21 &0.68 &0.65 &-0.15 &-0.18 &0.69 &0.67
\tabularnewline
\bottomrule
\end{tabular}
}
\end{table}

\textbf{Fairness Test.}
For collaborative fairness comparison, we only analyze our FDPDDL and the distributed framework using DSSGD, neglecting the centralised framework and standalone framework, because parties do not collaborate in the standalone framework, and parties cannot get access to the trained global model in the centralised framework, the global model is only available in the form of ``machine learning as a service" (MLaaS). 
Table~\ref{tbl:MNIST_fairness} and Table~\ref{tbl:SVHN_fairness} list the calculated fairness of the distributed framework and our FDPDDL on MNIST and SVHN datasets using CNN and MLP architectures, under settings of different sharing level and imbalanced data partition. Similarly, Table~\ref{tbl:AdultHospital_fairness} lists the fairness results on Adult and Hospital datasets. In particular, we omit the results for the same sharing level setting, 
as fairness is a less concerned problem in this setting. All the results are averaged over five random trails. As is evidenced by the high positive correlation coefficient, with all of them above 0.5, FDPDDL achieves reasonably good fairness, which confirms the intuition behind fairness: the party who is less private and has more training data delivers higher accuracy. In contrast, as evidenced in Table~\ref{tbl:MNIST_fairness}, Table~\ref{tbl:SVHN_fairness}, and Table~\ref{tbl:AdultHospital_fairness}, the distributed framework exhibits poor fairness with significantly lower values than that of FDPDDL in all cases, with even negative values in some cases, manifesting the lack of fairness in the distributed framework. This is because in the distributed framework, all the participating parties can derive similarly well local models no matter how much they contribute.

\begin{figure*}[!htp]
\centering
        \begin{subfigure}[ht]{0.235\textwidth}
                \includegraphics[width=1.02\linewidth,height=3.7cm]{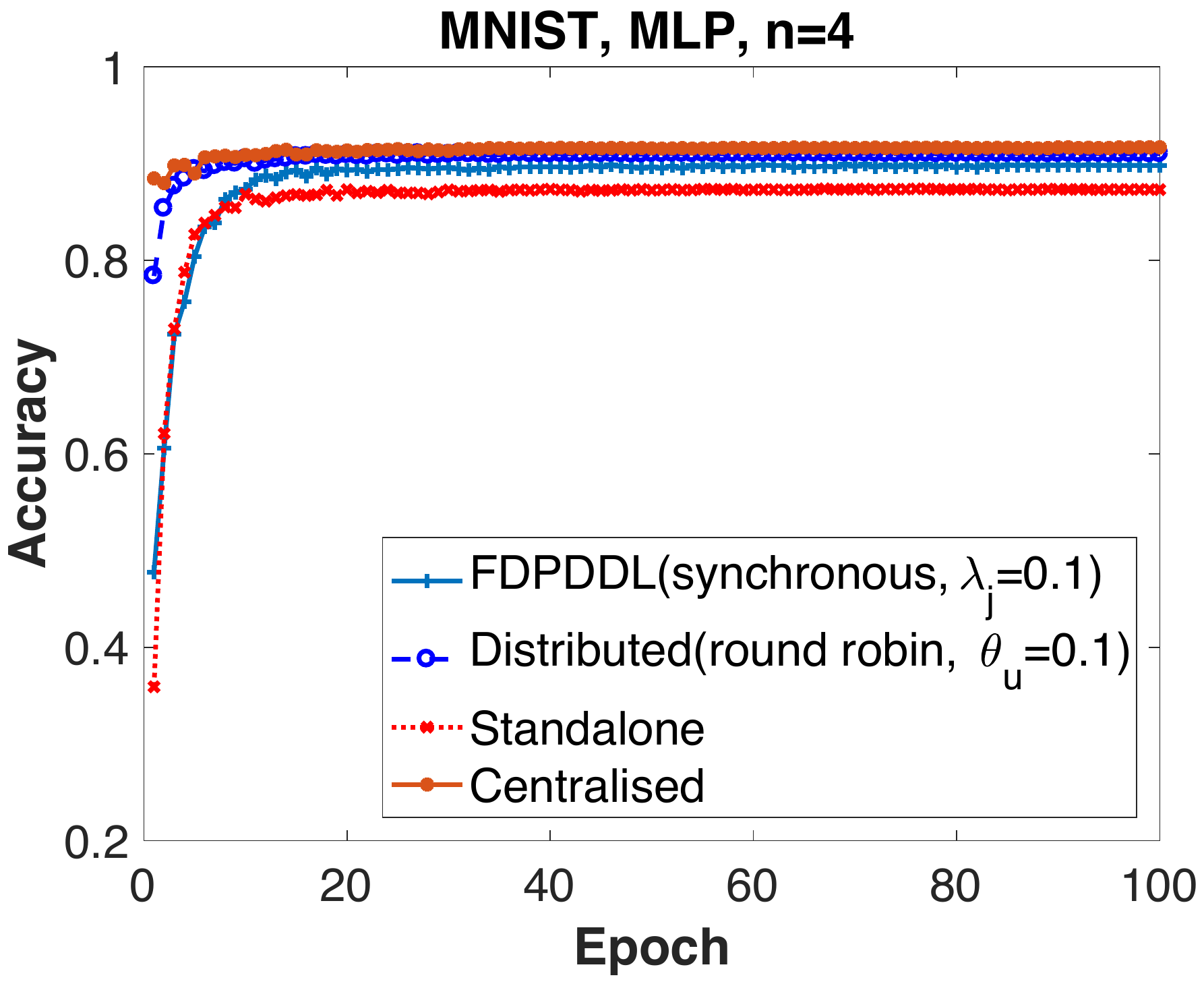}\label{fig:mnist_party4_epoch100_deep}
        \end{subfigure}
        \begin{subfigure}[ht]{0.235\textwidth}
                \includegraphics[width=1.02\linewidth,height=3.7cm]{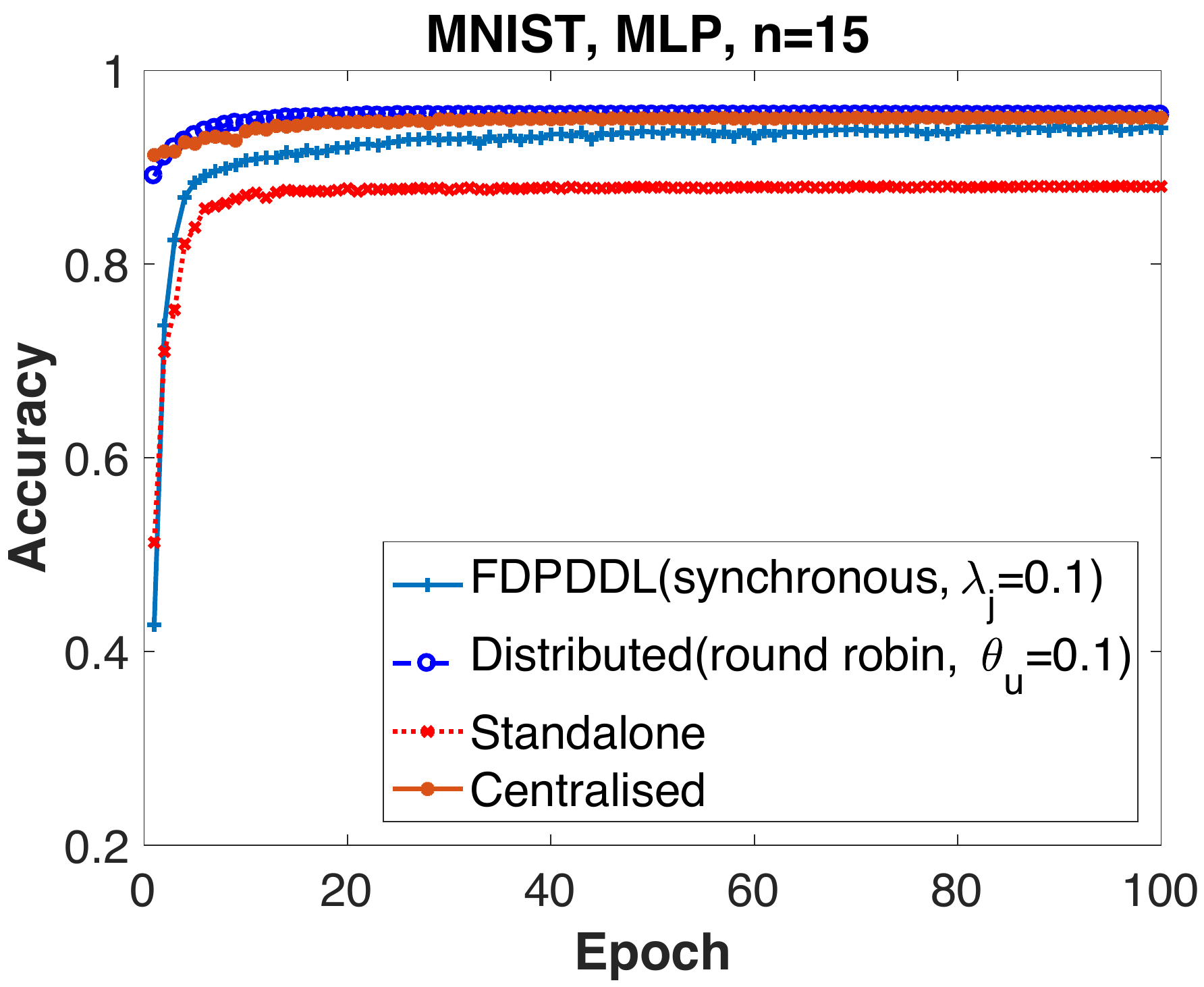}\label{fig:mnist_party15_epoch100_deep}
        \end{subfigure}
        \begin{subfigure}[ht]{0.235\textwidth}
                \includegraphics[width=1.02\linewidth,height=3.7cm]{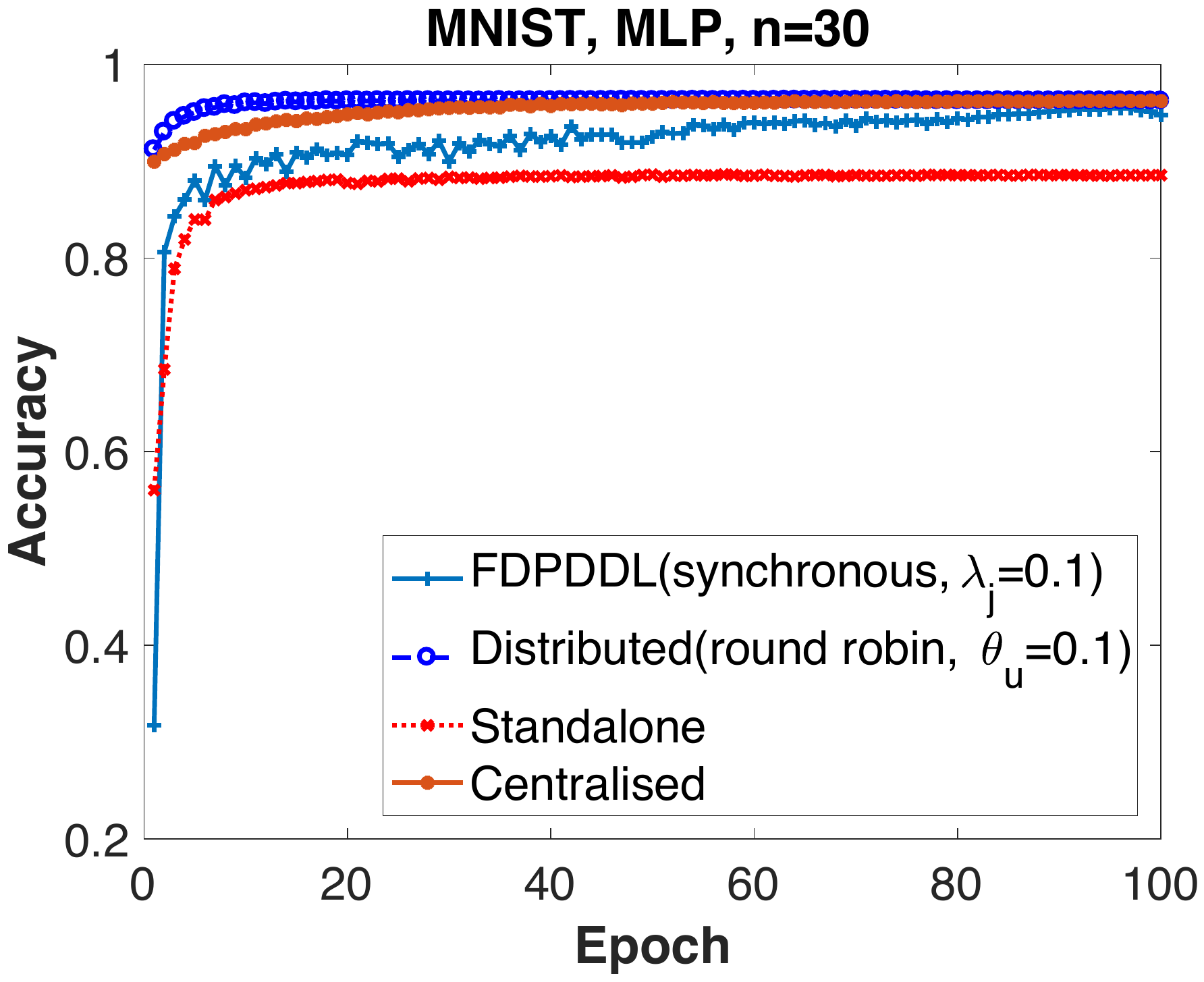}\label{fig:mnist_party30_epoch100_deep}
        \end{subfigure}
        \begin{subfigure}[ht]{0.235\textwidth}
                \includegraphics[width=1.02\linewidth,height=3.7cm]{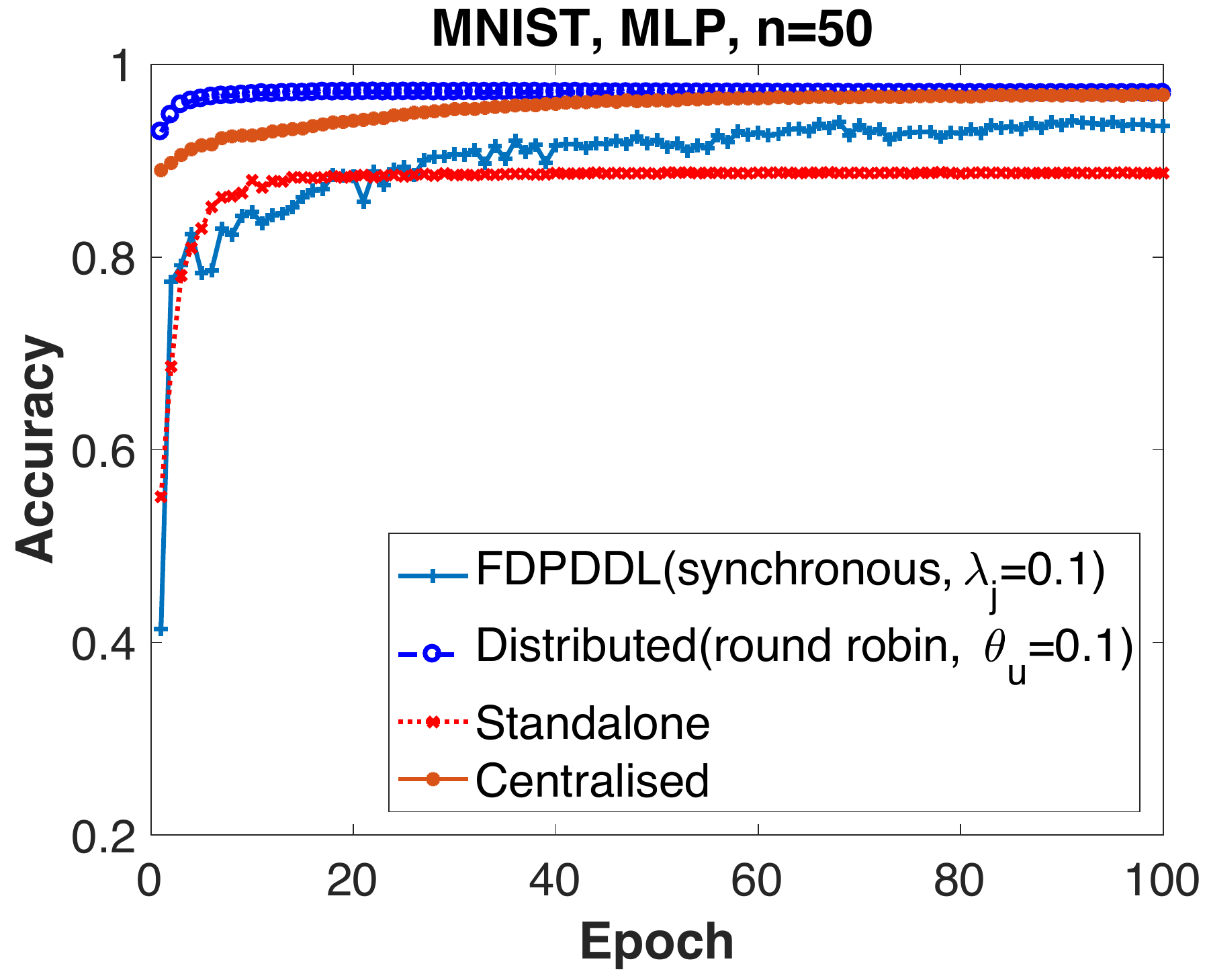}\label{fig:mnist_party50_epoch100_deep}
        \end{subfigure}
        \caption{MLP convergence on MNIST dataset for different frameworks and varying number ($n$) of parties.}
\label{fig:mnist_epoch100_deep}
\end{figure*}

\textbf{Learning Accuracy.}
For accuracy comparison, we implement FDPDDL using synchronous SGD protocol, and set the sharing level of each party to $0.1$ ($\lambda_j=0.1$). Similarly, for distributed framework, we implement DSSGD without differential privacy using round robin protocol, and set the upload rate $\theta_u$ to $0.1$. It is worth noting that we did not apply any privacy-preserving techniques to all the other three baseline frameworks in order to assess the impact of our FDPDDL on accuracy. For MNIST dataset, Fig.~\ref{fig:mnist_epoch100_deep} demonstrates that FDPDDL does not sacrifice much model utility when compared to the distributed or the centralised framework, meanwhile it delivers better accuracy than the standalone framework.

\begin{table*}[ht]
\caption{Maximum accuracy [\%] on MNIST under varying party number settings, achieved by \emph{Centralised}, \emph{Distributed} (DSSGD without DP, round robin, $\theta_u=10\%$), \emph{Standalone} and FDPDDL (see Section~\ref{sec:Setup} for details of the three settings) frameworks using MLP and CNN architectures. P-$k$ indicates there are $k$ parties participating in the learning process.}
\vspace{-1mm}
\label{tbl:MNIST_CNN}
\centering
\begin{tabularx}{\linewidth}{l*{8}{C{1}}}
\multirow{2}{*}{Framework} & \multicolumn{4}{c}{MLP} & \multicolumn{4}{c}{CNN}
\tabularnewline
\cmidrule(lr){2-5}\cmidrule(lr){6-9}
 & P4 & P15 & P30 & P50 & P4 & P15 & P30 & P50
\tabularnewline
\midrule
\textit{Centralised} 
& 91.68 & 95.17 & 96.28 & 96.85 
& 96.58 & 98.19 & 98.52 & 98.58
\tabularnewline
\textit{Distributed} 
& 91.67  & 95.17 & 96.33 & 97.35 
& 96.25  & 98.04 & 98.63 & 98.83
\tabularnewline
\textit{Standalone} 
& 87.39 & 88.06 & 88.64  & 88.80 
& 93.81 & 93.46 & 94.04 & 94.05
\tabularnewline
\textit{FDPDDL (same $\lambda_i$, same $|D_i|$)} 
& 88.44 & 92.18 &93.50 & 95.33 
& 94.34  & 96.89  & 97.62  & 97.67
\tabularnewline
\textit{FDPDDL (different $\lambda_i$, same $|D_i|$)}
& 89.84  & 92.82  & 93.50 & 95.37
& 94.62  & 96.15  & 97.78 & 98.05
\tabularnewline
\textit{FDPDDL (different $|D_i|$, same $\lambda_i$)}
& 88.92  & 91.44  & 92.69 & 95.02
& 94.39  & 96.46  & 96.92 & 97.47
\tabularnewline
\bottomrule
\end{tabularx}
\end{table*}

\begin{table*}[ht]
\caption{Maximum accuracy [\%] on SVHN under varying party number settings using MLP and CNN architectures.}
\label{tbl:SVHN_CNN}
\vspace{-1mm}
\centering
\begin{tabularx}{\linewidth}{l*{8}{C{1}}}
\multirow{2}{*}{Framework} & \multicolumn{4}{c}{MLP} & \multicolumn{4}{c}{CNN}
\tabularnewline
\cmidrule(lr){2-5}\cmidrule(lr){6-9}
 & P4 & P15 & P30 & P50 & P4 & P15 & P30 & P50
\tabularnewline
\midrule
\textit{Centralised}
& 75.40  & 83.08  & 85.77  & 87.15
& 90.50  & 91.88  & 93.42  & 95.44
\tabularnewline
\textit{Distributed}
&78.34   &85.49   &87.64 & 89.21
&91.78   &93.03   &95.75 & 96.19
\tabularnewline
\textit{Standalone}
& 57.85  & 58.77   & 57.90 & 59.18
& 80.24  & 80.74  & 81.29  & 81.60
\tabularnewline
\textit{FDPDDL (same $\lambda_i$, same $|D_i|$)}
& 67.74   & 76.55   &81.86 & 84.51
& 88.07   & 90.18   &90.74 & 92.83
\tabularnewline
\textit{FDPDDL (different $\lambda_i$, same $|D_i|$)}
&68.16   &76.67  &79.25 & 83.57
&88.91   &90.15  &91.29  & 93.18
\tabularnewline
\textit{FDPDDL (different $|D_i|$, same $\lambda_i$)}
&68.57   &74.15   & 80.37 & 83.34
&89.53   &90.03   & 92.13 & 93.82
\tabularnewline
\bottomrule
\end{tabularx}
\end{table*}

\begin{table*}[ht]
\caption{Maximum accuracy [\%] on Adult and Hospital under varying party number settings using MLP.}
\label{tbl:AdultHospital_MLP}
\vspace{-1mm}
\centering
\begin{tabularx}{\linewidth}{l*{8}{C{1}}}
\multirow{2}{*}{Framework} & \multicolumn{4}{c}{Adult} & \multicolumn{4}{c}{Hospital}
\tabularnewline
\cmidrule(lr){2-5}\cmidrule(lr){6-9}
 & P4 & P15 & P30 & P50 & P4 & P15 & P30 & P50
\tabularnewline
\midrule
\textit{Centralised}
& 80.69 & 81.54 &82.75 & 83.43
& 65.21 & 69.12 &74.50 & 76.21
\tabularnewline
\textit{Distributed}
&80.73   &81.89   &82.81 & 83.49
&65.58   &69.50   &74.73 & 77.12
\tabularnewline
\textit{Standalone}
& 78.49  & 78.50  & 78.52 & 78.54
& 53.51  & 53.71  & 53.89 & 53.95
\tabularnewline
\textit{FDPDDL (same $\lambda_i$, same $|D_i|$)}
& 79.08   &80.05   &81.16 & 82.21
& 63.21   &67.35   &72.38 & 74.55
\tabularnewline
\textit{FDPDDL (different $\lambda_i$, same $|D_i|$)}
&79.15   &80.08  &81.20 &82.28
&63.38   &67.42  &72.29 &74.62
\tabularnewline
\textit{FDPDDL (different $|D_i|$, same $\lambda_i$)}
&79.21   &80.17   & 81.25 & 82.39
&63.35   &67.58   & 72.41 & 74.58
\tabularnewline
\bottomrule
\end{tabularx}
\vspace{-3mm}
\end{table*}

Detailed accuracy comparison over varying participating parties ($n=\{4, 15, 30, 50\}$) can be found in Table~\ref{tbl:MNIST_CNN} for MNIST dataset, Table~\ref{tbl:SVHN_CNN} for SVHN dataset, and Table~\ref{tbl:AdultHospital_MLP} for Adult and Hospital datasets. As can be observed, the best test accuracy is reported by either the centralised framework or distributed framework using DSSGD without differential privacy, while the worst accuracy is given by the standalone framework (minimum utility, maximum privacy). In contrast, we observe that FDPDDL allows all parties to derive higher accuracies than that given by the standalone models trained on their local data alone, under all the investigated scenarios. This confirms the benefits brought to every party by the collaborative learning in our FDPDDL. Meanwhile, our FDPDDL also achieves comparable accuracy to the centralised framework and distributed framework using DSSGD without differential privacy, substantiating the competitive effectiveness of our decentralised framework.

Combining the above fairness results in Table~\ref{tbl:MNIST_fairness} Table~\ref{tbl:SVHN_fairness}, and Table~\ref{tbl:AdultHospital_fairness}, and accuracy results in Table~\ref{tbl:MNIST_CNN}, Table~\ref{tbl:SVHN_CNN} and Table~\ref{tbl:AdultHospital_MLP}, we conclude that \emph{FDPDDL achieves both fairness and privacy without severely harming accuracy}. This proves that our FDPDDL is a promising framework for effective, privacy-preserving and more importantly fair collaborative learning.

\textbf{Complexity Analysis}. Considering complexity, the main communication cost occurs when each party sends its differentially private samples or the selected differentially private gradients to the other $(n-1)$ parties, resulting in $(n-1)*L$ cost, where $n$ and $L$ are the number of parties and the average size of the released samples and gradients. It should be noted that parties do not share all their model updates with other parties, they selectively share model updates as per download request and their sharing levels, as explained in Section~\ref{subsec:focuses}. Therefore, we remark that our FDPDDL is more relevant to practical applications in horizontally federated learning (HFL) to businesses (H2B)~\cite{lyu2020threats}, such as biomedical or financial institutions where the number of parties $n$ is not too large, while the collaborative fairness is a more concerned problem. On the other hand, the main computation cost occurs at each party who needs to train a local DPGAN and local model to initialise local credibility and tokens during the first stage, and conduct local training and mutual evaluation of local credibility during the second stage. However, we remark that parties can train their DPGAN models and generate massive DPGAN samples offline, as parties are required to share their DPGAN samples only once during the first stage of initialisation, it does not affect the second stage of update, as shown in Fig.~\ref{fig:FDPDDL_flow}. For the update stage, all parties can individually update their local models in parallel. Moreover, using DP instead of encryption-based technique~\cite{lyu2020towards} during the second stage results in less communication cost. 

\subsection{Malicious Party Detection}
We further demonstrate how our framework can provide robustness to two specific malicious parties: ``free-riders'' and GAN attacker. 

\textbf{Robustness to ``free-riders''.}
In collaborative system, free-riders may pretend to be contributing by generating fake information to release to the requester. The main incentives for free-rider to submit fake information may include: (1) one party may not have any data to train a local model; (2) one party is too concerned about data privacy to release any information that may compromise privacy; (3) one party may not want to consume any local computation power to train any model. As demonstrated in Example.~\ref{exmp:malicious}, it is possible for a free-rider without any data or model to have access to the same global model. We simulate two possible strategies that such a free-rider party could exploit to achieve its goals.

\emph{Release random labels:} During the initialisation stage, the free-rider can release random labels for the received DPGAN samples. The initialisation stage allows each participant to evaluate the data quality of other participants before collaborative learning starts. If a participant does not have reasonable amount of training data to produce a decent model, it will perform poorly in the evaluation of DPGAN samples sent from other parties, thus other parties would assign low local credibility to this party to ensure fairness. More specifically, when the publisher receives the random labels from the free-rider, it will find that these random labels are not consistent with the majority voting, \ie $\frac{m_j}{u_i} \ll c_{th}$, then the free-rider will be reported as "malicious". If the majority party report one party as "malicious", then the blockchain will opt this party out in the future communications. Even though the free-rider might succeed in initialisation somehow, the credibility of the free-rider is significantly lower compared with the other honest parties, and the other parties will download less gradients from this free-rider.

\emph{Release random or carefully crafted gradients:} During the update stage, the free-rider may publish meaningless gradients such as random or carefully crafted gradients to pretend that it is contributing, but do not wish to be ``caught'' in cheating (keep stealthy). However, such meaningless gradients will further downgrade its local credibilities to all the other parties during the local credibility and tokens update stage (as described in Section~\ref{sec:credit_update}). Consequently, the free-rider will gradually lose its chance to earn more tokens as more and more parties downgrade its local credibility. The tokens of the free-rider will drain out faster and eventually be blocked out from the learning process when its tokens are used up. This can be automatically done by our reputation system through digital tokens and local credibility and the Blockchain protocol itself.

We simulate all the above malicious behaviors of the free-rider and track the collaboration process among parties. We notice that in most cases, the free-rider can be detected and excluded at the initialisation stage, and no free-riders can survive two stages.

\textbf{Robustness to GAN Attacker.}
We next discuss the robustness of our FDPDDL framework against GAN attacks~\cite{hitaj2017deep}. As argued in~\cite{hitaj2017deep}, GAN attacks can only succeed if the class distributions of the adversary and the victim party are Non-IID. Therefore, following the same setting as in GAN attack~\cite{hitaj2017deep} on MNIST dataset, we assume the victim parties own local data of class $\{0, 1, 2, 3, 4\}$ and the adversary has data of class $\{5, 6, 7, 8, 9\}$, that is, Non-IID class distribution between the adversary and victim parties. We confirmed empirically that our FDPDDL framework can successfully detect and isolate such kind of adversary. In particular, the initial local credibility of the adversary should be rated quite low by most parties during the initialisation stage, \ie $\frac{m_j}{u_i} \ll c_{th}$. Therefore, the GAN adversary can be detected and excluded mostly at the initialisation stage. Even though the GAN adversary can somehow survive the initialisation stage, in the subsequent update stage, the accumulated local credibility of the adversary is rated even lower when it iteratively publishes false and meaningless gradients. Eventually, the GAN adversary can be successfully detected and isolated by our FDPDDL when it is agreed as "malicious" through the blockchain consensus.

Recall that for GAN attack, the adversary needs to learn an extra GAN network during the collaborative learning process and this requires expensive computation. This inevitably results in suspicious longer training time than the honest parties. Therefore, the response time characteristic can be further incorporated into our credibility mechanism to greatly reduce the chance of privacy leakage. If one party does not respond within a reasonable amount of time, other parties should anyway assume that this party is down and discard its submission at the current round.
Moreover, even if we assume that the malicious party somehow manages to circumvent the above mechanism, there is only little harm because the subsequent gradients submitted by the malicious party will not be rated highly by the honest parties, and its local credibility will get progressively reduced.

\section{Discussion}
\textbf{Attacking Fairness in FDPDDL.} 
Here, we discuss some possible strategies that can be exploited by a free-rider or a party with very little data to deceive other parties and gain unfairly from our FDPDDL. For a free-rider owning no local data, it can manually label the received DPGAN examples as its local data and publish the labels to cheat the initialisation stage and the subsequent update stage. But this can be extremely expensive and practically unachievable for more complex collaborative learning tasks.

Similarly, for a malicious party having very little local data, it can make use of the received DPGAN examples to first train a good representation extractor via unsupervised representation learning (\eg autoencoder), then build a good local classification model on top of the representation extractor using its labelled local data. Implementing this may improve the local model quality of this party and increase its credibility, thus seems to increase the risk of privacy leakage towards this party. However, we remark that the privacy of honest parties in FDPDDL should not be affected under such malicious behaviors, as secured by our majority voting mechanism used in the initialisation stage. That is, local improvement of a malicious party does not guarantee its high credibility as long as it conforms with the majority honest parties, otherwise even decreases its credibility.

In practice, there may also exist honesty challenge imposed by a group of malicious parties. For instance, some malicious parties might collude with each other to downgrade the credibility of an honest party and to block it out from the learning process. Or they can upgrade each other's credibilities. However, as far as honesty is concerned, FDPDDL can always be able to detect such malicious behaviors as long as a majority of the participants are honest.

\textbf{Collaborative Learning and GANs.} We apply data augmentation to expand local data size to help DPGAN generate reliable samples for local credibility initialisation, and facilitate the implementation of DPSGD in collaborative learning, as larger amount of local data allows for more iterations in training a DPGAN or a differentially private local model within a moderate privacy budget. One natural question is that, if we can generate infinite examples using GANs, why do we still need collaborative learning? This is because GANs can only learn the local data distribution, which means the examples generated by GANs are restricted to local data distribution, while collaborative learning is specially designed to break such local restrictions through benefiting from global collaboration~\cite{shokri2015privacy}. Note that using DPGAN and DPSGD in FDPDDL instead of the standard GAN and SGD not only preserves the training data privacy, but also preserves the privacy of the augmented data~\cite{xie2018differentially}.

\section{Conclusion and Future Work}
\label{sec:Conclusion}
This work is the first step in bringing fairness and privacy to democratise and protect AI, hence providing better incentive for more parties to collaborate. Our proposed Fair and Differentially Private Decentralised Deep Learning (FDPDDL) framework demonstrates the following properties: (1) it inherently solves the single-point-of-failure problem existing in all server-based frameworks; (2) it achieves high fairness by creating a reputation system through digital tokens and local credibility, which considers the relative contributions of all the parties through two novel algorithms: local credibility and tokens initialisation, and local credibility and tokens update; (3) it provides a viable solution to detect even isolate the "non-credible" party both before the collaborative learning process starts, and during the collaborative learning process, in this way, our scheme provides robustness to the malicious "free-riders" and GAN attacker; (4) \emph{Differentially Private GAN} and \emph{Differentially Private SGD} are used to guarantee local privacy of each party. The experimental results on benchmark datasets in three realistic settings demonstrate that FDPDDL framework consistently outperforms the standalone framework and achieves comparable accuracy to the centralised framework and distributed framework without differential privacy, confirming the applicability of FDPDDL. We believe our findings could be inspiring for the follow-up research in the decentralized learning, especially we initiate a new field of collaborative fairness in such an environment. For future work, we would like to consider more advanced model architectures, different attacks in distributed/decentralised learning and Non-IID data.  

\section*{Acknowledgment}
This work is partially supported by Faculty of Information Technology, Monash University; and an IBM PhD Fellowship. 

\bibliographystyle{IEEEtran}
\bibliography{biblio}
\end{sloppypar} 

\vskip -1\baselineskip plus -1fil
\begin{IEEEbiography}[{\includegraphics[width=1in,height=1in,clip,keepaspectratio]{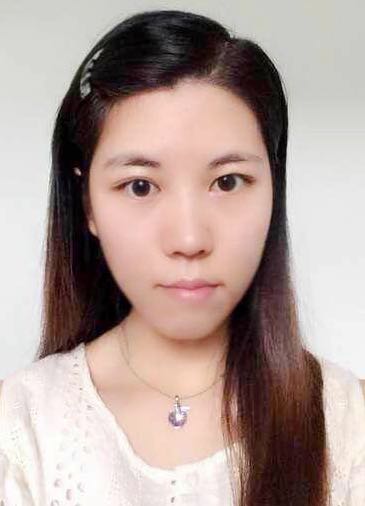}}]{Lingjuan Lyu}
(IEEE M'18) is currently a Research Fellow with The Department of Computer Science, NUS. She received Ph.D. degree from the University of Melbourne. Her current research interests span machine learning, privacy, fairness, and edge intelligence. 
\end{IEEEbiography}

\vskip -3.5\baselineskip plus -1fil

\begin{IEEEbiography}[{\includegraphics[width=1in,height=1in,clip,keepaspectratio]{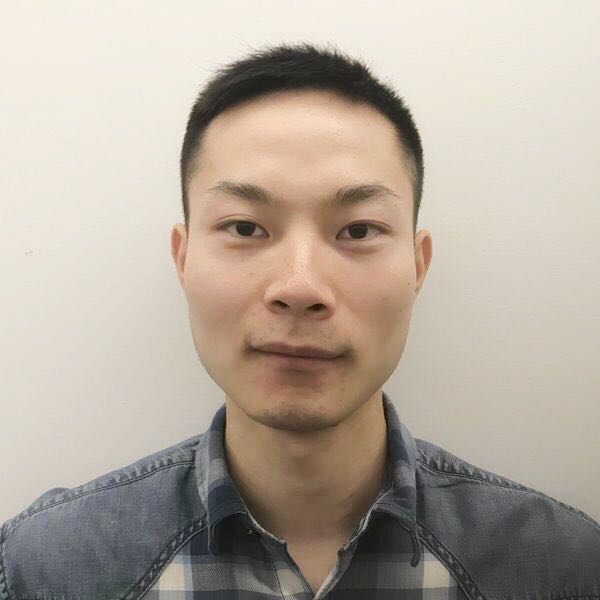}}]{Yitong Li}
is currently a Ph.D student in School of Computing and Information Systems, the University of Melbourne. He received B.S. degree from Shanghai Jiao Tong University. His research interests cover privacy and adversarial learning with NLP applications. 
\end{IEEEbiography}

\vskip -3.5\baselineskip plus -1fil

\begin{IEEEbiography}[{\includegraphics[width=1in,height=1in,clip,keepaspectratio]{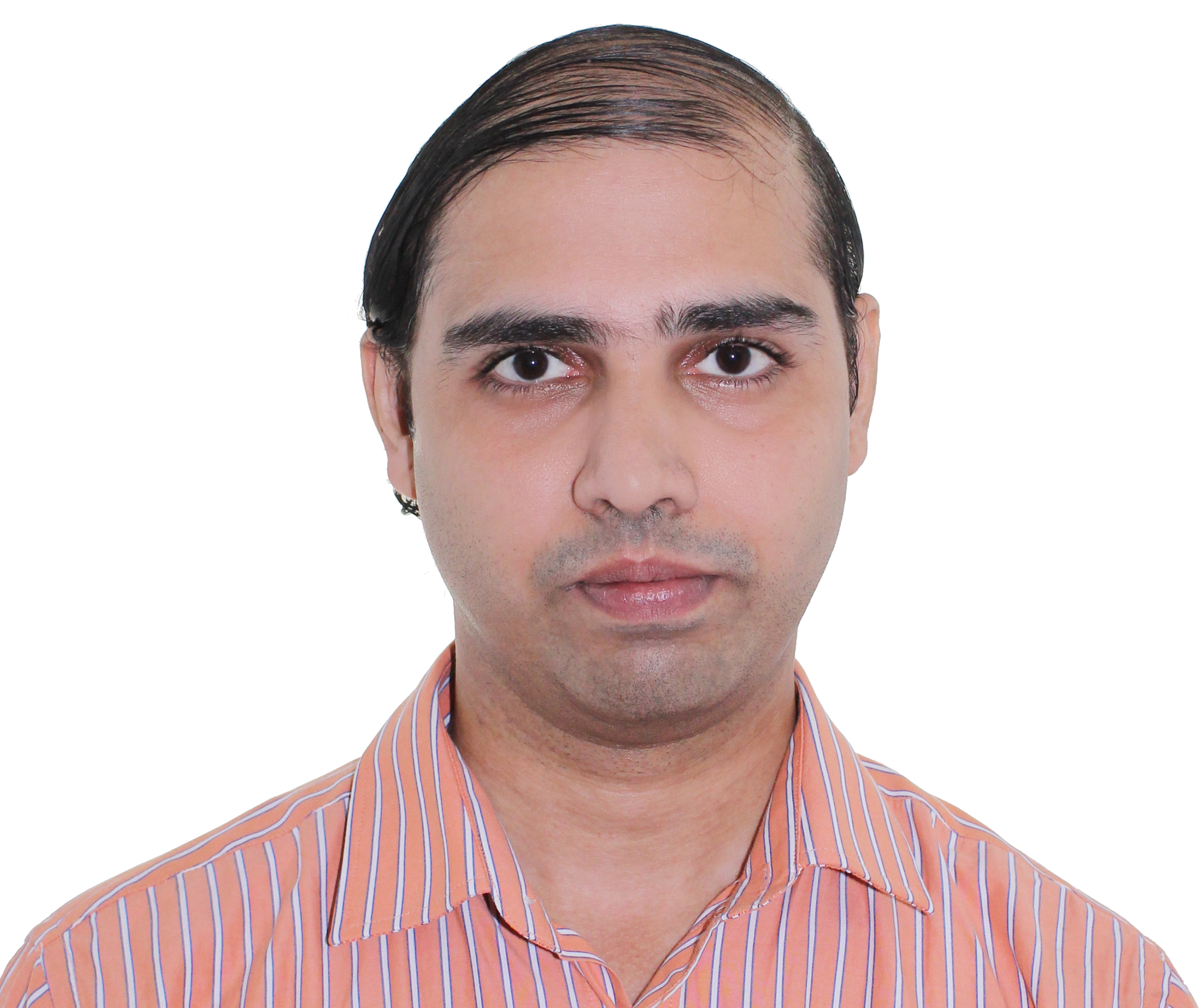}}]{Karthik Nandakumar}
(IEEE SM'02) is a Research Staff Member at IBM Research, Singapore. Prior to joining IBM in 2014, he was a Scientist at Institute for Infocomm Research, A*STAR, Singapore for more than six years. He received his B.E. degree (2002) from Anna University, Chennai, India, M.S. degrees in Computer Science (2005) and Statistics (2007), and Ph.D. degree in Computer Science (2008) from Michigan State University, and M.Sc. degree in Management of Technology (2012) from National University of Singapore. His research interests include computer vision, statistical pattern recognition, biometric authentication, image processing, machine learning and blockchain.
\end{IEEEbiography}

\vskip -2.5\baselineskip plus -1fil

\begin{IEEEbiography}[{\includegraphics[width=1in,height=1in,clip,keepaspectratio]{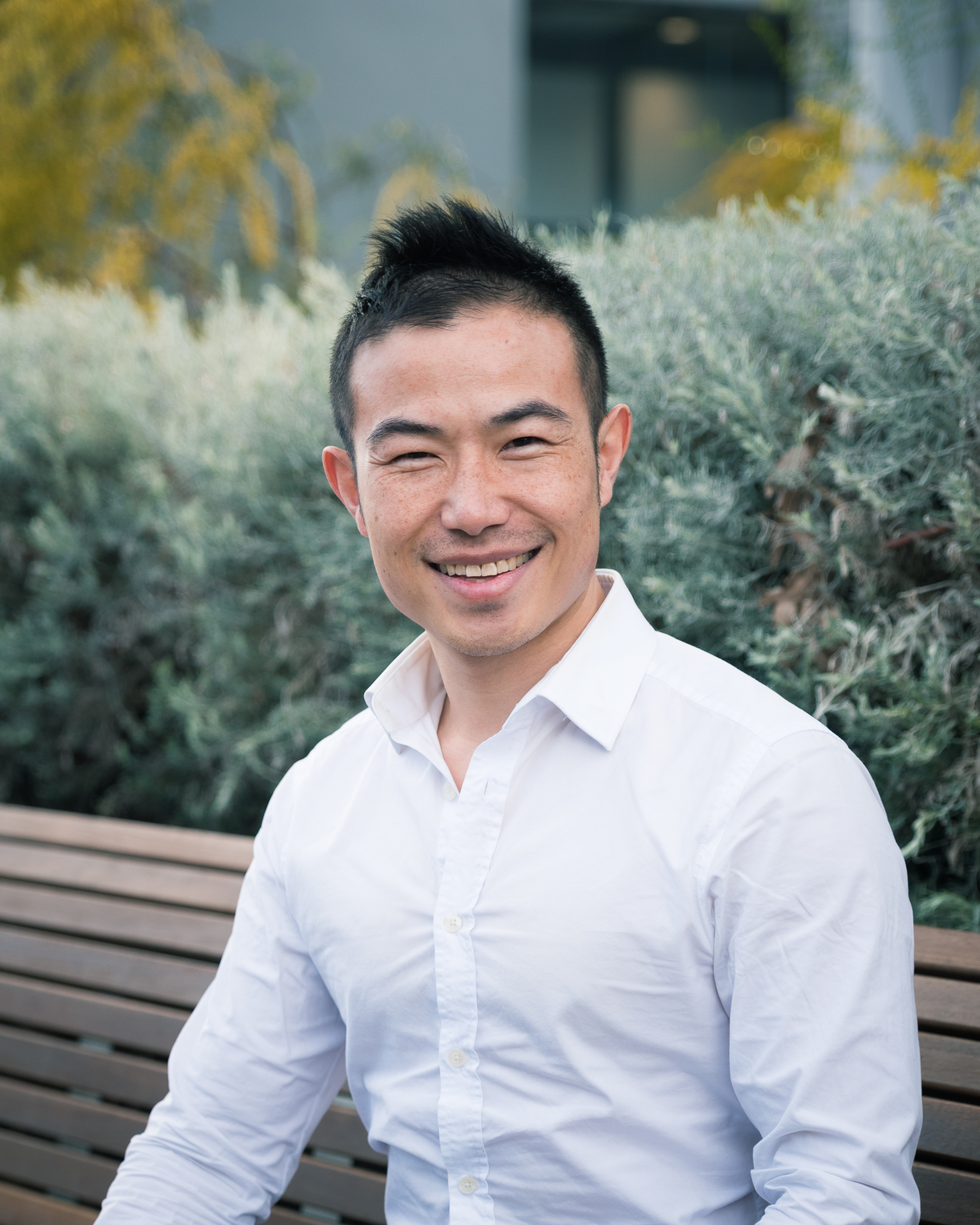}}]{Jiangshan Yu}
received the Ph.D. degree from the University of Birmingham (UK) in 2016. He is currently Associate Director (Research) at Monash Blockchain Technology Centre at Monash University, Australia. Previously, he was a research associate at SnT, University of Luxembourg (LU). The focus of his research has been on design and analysis of cryptographic protocols, cryptographic key management, blockchain consensus, and ledger-based applications. 
\end{IEEEbiography}

\vskip -3.5\baselineskip plus -1fil

\begin{IEEEbiography}[{\includegraphics[width=1in,height=1in,clip,keepaspectratio]{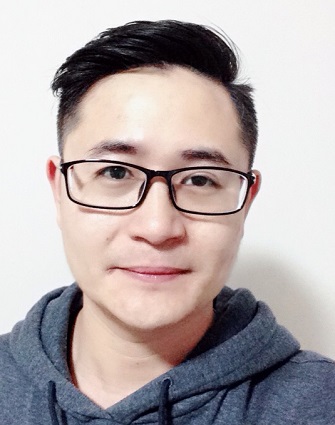}}]{Xingjun Ma}
is a lecturer in School of Information Technology, Deakin University, and also an honorary fellow in School of Computing and Information Systems, The University of Melbourne. He received Ph.D., M.E., B.E. degrees from the University of Melbourne, Tsinghua University and Jilin University respectively. He works in the areas of adversarial machine learning and robust optimisation.
\end{IEEEbiography}
\end{document}